\documentclass[a4j, 11pt]{article}
\usepackage[top=20truemm,bottom=20truemm,left=20truemm,right=20truemm]{geometry}
\usepackage{physics}
\usepackage[dvipdfmx]{graphicx}
\usepackage{amsmath, amssymb}
\usepackage[dvipsnames,x11names]{xcolor}
\usepackage{hyperref}
\usepackage{subcaption}
\hypersetup{colorlinks=true, citecolor=blue, urlcolor=blue, linkcolor=blue}
\usepackage{cite}
\usepackage{siunitx}
\usepackage{authblk} 
\usepackage{pgf}
\usepackage{ulem}

\title{On the Galactic radio signal from stimulated decay of axion dark matter} 

\author{P.~S.~Bhupal Dev}
\author{Francesc Ferrer}
\author{Takuya Okawa\thanks{o.takuya@wustl.edu}}

\affil{\it Department of Physics and McDonnell Center for the Space Sciences,
\\ 
\it Washington University, Saint Louis, MO 63130, USA}

\def\atag{\refstepcounter{equation}\tag{\arabic{equation}}}
\begin{document}
\date{} 
\maketitle

\begin{abstract} 
	We study the full-sky distribution of the radio emission from the 
	stimulated decay of axions which are assumed to compose the dark 
	matter in the Galaxy. Besides the constant extragalactic and CMB 
	components, the decays are stimulated 
	by a Galactic radio emission with a spatial distribution that we 
	empirically determine from observations. We compare the diffuse
	emission to the counterimages of the brightest supernov\ae\
	remnants, and take into account the effects of free-free 
	absorption. We show that, if the dark matter halo is described by a
	cuspy NFW profile, the expected signal from the Galactic center 
	is the strongest. Interestingly, the emission from the Galactic 
	anti-center provides competitive constraints that do not depend
	on assumptions on the uncertain dark matter density in the inner
	region. Furthermore, the anti-center of the Galaxy is 
	the brightest spot if the Galactic dark matter density follows a 
	cored profile. The expected signal from stimulated decays of 
	axions of mass $m _{a} \sim 10 ^{-6} \si{\ eV}$
	is within reach of the Square Kilometer Array for an axion-photon 
	coupling $g _{a\gamma} \gtrsim (2-3) \times 10 ^{-11} \si{\ GeV} ^{-1}$. 

\end{abstract}

\section{Introduction}
\indent
The axion was originally introduced to address the strong CP problem in quantum chromodynamics (QCD)~\cite{Peccei:1977hh,Weinberg:1977ma,Wilczek:1977pj}. The QCD axion emerges as a pseudo-Nambu-Goldstone boson (pNGB) when the $U(1)_{\rm PQ}$ symmetry is broken. It can also be a viable cold dark matter candidate~\cite{Preskill:1982cy,Abbott:1982af,Dine:1982ah, Ipser:1983mw}. These considerations have been extended to axion-like particles (ALPs), i.e.~pseudoscalar particles that could generically appear as pNGBs in theories with a spontaneously broken global $U(1)$ symmetry, which are ubiquitous in string-inspired beyond the Standard Model (SM) constructions~\cite{Svrcek:2006yi, Arvanitaki:2009fg}. ALPs cover a much broader mass-coupling range than the original QCD axion~\cite{DiLuzio:2020wdo}, and have inspired many novel laboratory and astrophysical searches~\cite{Irastorza:2018dyq, Choi:2020rgn}.

A common characteristic of the QCD axion, as well as of ALPs, is that their couplings to the SM particles are suppressed by inverse powers of the $U(1)$ symmetry breaking scale, $f_a$. In this paper, we will only consider the ALP coupling to photons (or electromagnetic fields) which can be parametrized by the following Lagrangian:  
\begin{equation}
    -\mathcal{L}=\frac{1}{2} m_{a}^{2} a^{2}-\frac{g_{a \gamma} }{4} a F_{\mu \nu} \tilde{F}^{\mu \nu}  = \frac{1}{2} m_{a}^{2} a^{2}+ g_{a \gamma} a \mathbf{E} \cdot \mathbf{B}  ,
    \label{eq:lag}
\end{equation}
where $F _{\mu\nu}$ is the electromagnetic field strength tensor, 
$\tilde{F} ^{\mu\nu}$ is its dual, $a$ is the axion field, and 
$\mathbf{E}$ and $\mathbf{B}$ are the electric and magnetic fields,
respectively. The effective axion-photon coupling $g_{a\gamma}$ has the dimension of inverse mass; specifically, for the QCD axion, $g_{a\gamma}=\alpha C_{a\gamma}/(2\pi f_a)$, where $\alpha$ is the electromagnetic coupling strength and $C_{a\gamma}$ is a model-dependent dimensionless quantity typically of order unity; e.g. in the Kim-Shifman-Vainshtein-Zakharov (KSVZ)~\cite{Kim:1979if,Shifman:1979if} model, $C_{a\gamma}=-0.97$, and in the Dine-Fischler-Srednicki-Zhitnitsky (DFSZ)~\cite{Zhitnitsky:1980tq,Dine:1981rt}  and all grand-unified axion models, $C_{a\gamma}=0.36$.  Moreover, the QCD axion mass satisfies the relation $m_af_a\approx m_\pi f_\pi$, where $m_\pi$ and $f_\pi$ are the pion mass and pion decay constant, respectively. Including QED and NNLO corrections in chiral perturbation theory leads to the relation~\cite{Gorghetto:2018ocs}
\begin{align}
    m_a = 5.691(51)\mu {\rm eV}\left(\frac{10^{12}~{\rm GeV}}{f_a}\right) ,
\end{align}
thus restricting $m_a$ and $g_{a\gamma}$ to a narrow band. However, in general, the ALP mass $m_a$ and coupling $g_{a\gamma}$ are treated as independent free parameters, and the experimental constraints are often quoted in the $(m_a,g_{a\gamma})$ plane. 

As alluded to above, QCD axions or ALPs may also account for the dark matter in the Universe. They could be produced in the early Universe by several mechanisms~\cite{Sikivie:2006ni, Marsh:2015xka}, such as vacuum misalignment~\cite{Preskill:1982cy,Abbott:1982af,Dine:1982ah}, thermal production~\cite{Turner:1986tb, Covi:2001nw, Brandenburg:2004du, Salvio:2013iaa}, collapse of cosmic strings and domain walls~\cite{Davis:1986xc, Harari:1987ht, Lyth:1991bb, Hiramatsu:2012gg, Buschmann:2021sdq}, decay of heavier particles (e.g.~moduli or inflaton coupling to axions)~\cite{Acharya:2010zx, Cicoli:2012aq, Higaki:2012ar, Conlon:2013isa, Higaki:2013lra, Baer:2022fou, Cicoli:2023opf} and Hawking radiation from primordial black holes~\cite{Schiavone:2021imu, Mazde:2022sdx, Li:2022xqh}. For instance, in the misalignment mechanism, 
the induction of nonzero QCD axion mass by non-perturbative QCD 
instanton effects at around the QCD phase transition $T \sim \Lambda_{\rm QCD}$
triggers oscillations of the axion field that could yield 
the required amount of the dark matter in the Universe 
for QCD axions in mass range $m_a \sim \mu$eV, 
which is our main focus here.

Multiple lines of reasoning have been considered in the literature
to constrain the parameter space of ALPs in the $\mu$eV mass region~\cite{AxionLimits}. 
The coupling to electromagnetic fields causes the conversion of axions into photons in the presence of an external magnetic field~\cite{Sikivie:1983ip}. Haloscopes such as the
Axion Dark Matter eXperiment 
(ADMX)~\cite{
ADMX:2018gho,
ADMX:2019uok,
ADMX:2021nhd,ADMX:2021mio} use a strong magnetic field applied 
inside a resonant cavity to find evidence of the conversion of axions to photons in the Galactic dark
matter halo. Assuming that axions compose all the dark matter, the absence of 
a signal at ADMX puts an upper bound on the axion-photon coupling $g_{a\gamma} \lesssim 3\times 10^{-16}$ GeV$^{-1}$ (at 90\% confidence level (C.L.)) in the $\mu$eV mass range and already excludes both KSVZ and DFSZ models over a narrow mass range of $2.8-4.2 \si{\ \mu eV}$~\cite{ADMX:2019uok, ADMX:2021nhd}. 

The same coupling also induces the production of axions from photons inside stellar objects. Helioscopes like the CERN Axion Solar Telescope
(CAST)~\cite{
CAST:2017uph} searched for axions from the Sun, and placed a bound on  
$g_{a \gamma}<6.63 \times 10^{-11} \ \mathrm{GeV}^{-1}$~(at 95\% C.L.) for 
axions with mass $m_{a} \lesssim 0.02 \ \mathrm{eV}$, which remains one of the most stringent laboratory constraints on ALPs in a wide mass range.  

Axions can also be searched for by considering astrophysical processes. 
Axion-photon conversion inside stellar cores affects the ratio of 
horizontal branch stars to red giants ($R$ parameter). An analysis of 39 Galactic globular 
clusters gave $g_{a \gamma}<6.54 \times 10^{-11}\ \mathrm{GeV}^{-1}\
(95 \%$ C.L.$)$ for $m _{a} \lesssim 100 \si{\ keV}$~\cite{Ayala:2014pea}. This was recently updated to $g_{a \gamma}<4.7 \times 10^{-11}\ \mathrm{GeV}^{-1}\
(95 \%$ C.L.$)$ from the observed ratio of asymptotic giant branch to horizontal branch stars ($R_2$ parameter) using a semiconvective mixing scheme~\cite{Dolan:2022kul}. Using the predictive mixing convective boundary scheme, as favored by asteroseismological evidence, improves the bound to $g_{a \gamma}<3.4 \times 10^{-11}\ \mathrm{GeV}^{-1}$~\cite{Dolan:2022kul}. 

Similarly, axion DM may efficiently convert to photons in the magnetospheres of neutron stars, producing nearly monochromatic radio emission. Using archival Green Bank Telescope data collected in a survey of the Galactic Center in the C-Band by the Breakthrough Listen project, constraints on $g_{a\gamma}$ down to the level of $10^{-11}~{\rm GeV}^{-1}$ was set for axion DM masses between 15 and 35 $\mu$eV~\cite{Foster:2022fxn}. Another study that does not rely on axions being the DM uses the fact that axions can be copiously produced in localized regions (polar caps) of neutron star magnetospheres from the spacetime oscillations of ${\bf E}\cdot {\bf B}$, and their resonant conversion into photons can generate a large broadband contribution to the neutron star’s intrinsic radio flux~\cite{Prabhu:2021zve}. Comparing observations of 27 nearby pulsars to predictions from sophisticated particle-in-cell simulations, an upper limit of $g_{a\gamma}\lesssim (2-10)\times 10^{-12}\ \mathrm{GeV}^{-1}\
(95 \%$ C.L.$)$ was obtained for $10^{-9}\lesssim m_a/{\rm eV}\lesssim 10^{-5}$~\cite{Noordhuis:2022ljw}.

Given the current constraints, QCD axions with a mass $\sim \mu$eV interact with the electromagnetic field very weakly. Thus, their spontaneous decay rate to two photons in vacuum is extremely small: 
\begin{align}
    \tau_a = \frac{64 \pi}{g_{a\gamma}^2 m_a^3 } \simeq 10^{43} \si{\ s}\left(\frac{10^{-10}\si{\ GeV^{-1}}}{g_{a\gamma}}\right)^2 \left(\frac{1~\mu\si{eV}}{m_a}  \right)^3  .
    \label{eq:vdecay}
\end{align}
This tiny decay rate makes it difficult to search QCD axions by photon signals from their spontaneous decay.\footnote{This is unlike other decaying DM candidates, like keV sterile neutrinos~\cite{Drewes:2016upu},  or heavy decaying  DM~\cite{Ibarra:2013cra}, where the photon signal constitutes one of the main probes.} However, the decay rate can be enhanced (by several orders of magnitude) in the presence of ambient photons  since electromagnetic fields with an  energy equivalent to half of ALP mass can stimulate decays of ALPs into two photons, which are emitted back-to-back in the rest frame of ALPs~\cite{Arza:2019nta}.
Thus, the incident wave of photons is amplified in the forward direction and reflected
in the opposite direction~\cite{Arza:2019nta, Arza:2021nec, Gong:2023ilg, Arza:2023rcs}. This phenomenon could generate an observable amplification  of radio signals from different astrophysical targets, such as dwarf spheroidal Galaxies, the Galactic Center and halo, and Galaxy clusters~\cite{Caputo:2018ljp,Caputo:2018vmy}. We can also observe photons produced from stimulated decay of ALPs in the exact opposite direction (antipodal point) to a bright astrophysical source in the rest frame of the axion. This counterimage, dubbed as {\it gegenschein}, has been studied for bright radio sources,  such as Cygnus A~\cite{Ghosh:2020hgd} and supernova remnants (SNRs)~\cite{Buen-Abad:2021qvj, Sun:2021oqp}. This {\it gegenschein} signal, obtained by integrating over all the axion decay in a DM column oriented along the line of sight, can in principle be detected using  powerful radio telescopes, like the current Five-hundred-meter Aperture Spherical radio Telescope (FAST)~\cite{Nan:2011um} or the future Square Kilometer Array (SKA)~\cite{Braun:2019gdo}. It will provide yet another probe of the axion-photon coupling $g _{a\gamma}$ in the $\mu$eV ALP mass range, which  falls right in the frequency range of operation of the radio telescopes: 
\begin{align}
    \nu_a=\frac{m_a}{4\pi}\simeq 0.1~{\rm GHz}~\left(\frac{m_a}{1~\mu{\rm eV}}\right)  .
    \label{eq:radio}
\end{align}

In this study, we carefully scrutinize the radio signals from 
stimulated decays of axions and include a number of effects previously not considered. 
We take into account the photon absorption due to electrons in the Galaxy
that could potentially affect the observed signal. We construct an all-sky map of the signal-to-noise ratio at different radio frequencies. In addition to the three brightest SNRs considered previously, we include a fourth SNR S147 (whose counterimage will be close to the Galactic Center), and also include the effect of SNR parameter uncertainties on the signal. The effect of DM density profile (cuspy versus cored) on the radio signal is also studied. We find that at high frequencies ($>$ GHz), the strongest signal is from the Galactic Center for both cuspy and cored profiles, whereas at low frequencies, the strongest signal is from either Galactic Center or Anti-center, depending on whether the density profile is cuspy or cored at the Center. Similarly, among the four point sources considered, S147 gives us the best limit for the cuspy profile. The sensitivities obtained here are comparable to the existing limits, and those from the Galactic Anti-center are especially robust against astrophysical uncertainties. Finally, we also quantify the potential effects of axion miniclusters and mass segregation due to dynamical friction on the radio signal, and find that their effect on the signal flux is negligible (at most 0.3\%).   

The rest of the paper is structured as follows: 
in Section~\ref{sec: stimulated decay of axion} we review the calculation of
the photon flux from the stimulated decay of axions (with additional details
reported in Appendix~\ref{sec:axion_decay_details}) and we discuss some factors overlooked before that may affect the signal observed at the Earth. 
Section~\ref{sec: the milky way} describes the Galactic environment 
and the characteristics of the SNRs considered here. 
The specifications of the relevant telescopes used in this analysis are summarized in 
Section~\ref{sec: telescopes}. The results are discussed 
in Section~\ref{sec: results}. Our conclusions are given  in 
Section~\ref{sec: conclusions}.

\section{Radio signal from the stimulated decay of axion dark matter}
\label{sec: stimulated decay of axion}

Let us consider a bright radio source such as an SNR in the
galaxy or a powerful extragalactic
source like Cygnus A. The beam of radio photons may
induce the stimulated decay of axions on its way through the Galactic DM halo to the Earth and 
beyond. The {\it gegenschein} signal~\cite{Ghosh:2020hgd} is generated along the continuation of the line of sight to the source, but to the direction opposite to the source. This could be a clean signal if there is no other bright source in the opposite direction of the original bright source in the sky. In contrast, the stimulated signal generated along the line of sight in the same direction as the original source will be difficult to isolate from the bright continuum background of the source itself. 

The
observed photon flux 
has contributions from the decay of axions
at a distance $r$ away from the Earth that occurred at $t - r/c$, and
was stimulated by a photon that had passed the current location of the Earth
at time $t - 2r/c$, where $c$ is the speed of light in vacuum.\footnote{We set $c=1$ elsewhere in the text, but keep it here for clarity.} If the radio source was significantly brighter in the 
past, as expected for some SNRs~\cite{Buen-Abad:2021qvj, Sun:2021oqp},
it could generate a striking bright image in the opposite direction, which is the {\it gegenschein} signal, whose frequency is set by the axion mass.

The flux density of photons from stimulated decays of axions averaged 
over the photon bandwidth $\Delta \nu$ is given by (see Appendix~\ref{sec:axion_decay_details} for the detailed derivation) 
\begin{align}
    \label{eq:flux_stimulated}
    S _{\nu} = \frac{\Gamma_a}{4\pi\Delta\nu} \int\dd x \int \dd \Omega\ \rho _{a}(x,\Omega) e ^{-\tau(\nu_a,x,\Omega)} \left(f_{\gamma}(x,\Omega,t)+\tilde{f}_{\gamma}(x, \Omega, t) \right), \atag
\end{align}
where $\Gamma_a=\tau_a^{-1}$ is the spontaneous axion decay rate [cf.~Eq.~\eqref{eq:vdecay}], $\rho_a$ is the axion mass density along the line of sight (which depends on the DM density profile), $\tau$ is the optical depth (see below), and $\nu_a$ is the frequency at which the signal peaks for a given axion mass $m_a$ [cf.~Eq.~\eqref{eq:radio}]. Here, we integrate the solid angle $\Omega$ over the field of view of the radio
telescope under consideration. The distribution functions $f _{\gamma}$ and $\tilde{f _{\gamma}}$ 
correspond to photons moving towards and away from the Earth, respectively,\footnote{Our flux expression differs from Eq.~(3.1) in Ref.~\cite{Caputo:2018vmy} in the last term where we take into account the fact that the flux of
photons traveling towards and away from the Earth may not necessarily be the same.} and 
their time dependence arises because of the time-dependent radio photons from point sources like SNRs. The radio photons from continuum sources, like the Galactic, extragalactic, and CMB emissions are treated as steady emissions (independent of time).

The exponential factor in Eq.~\eqref{eq:flux_stimulated} accounts for the
fact that a fraction of the photons from axion decays will be absorbed 
before they can reach the Earth. This is mainly due to free-free absorption,
which we model as a thermal plasma of ionized hydrogen uniformly mixed with the synchrotron-emitting relativistic gas. The absorption coefficient is given by~\cite{singer1961cosmic,gregory1974nature}
\begin{align}
    \label{eqn:absorption coefficient}
    \kappa(\nu,\mathbf{x})= (9.8 \times 10^{-3}~\mathrm{cm}^{-1}) [n_e(\mathbf{x})]^2 [T_e(\mathbf{x})]^{-3 / 2} \nu^{-2}\left[19.8+\ln \left(\frac{[T_e(\mathbf{x})]^{3 / 2}}{\nu}\right)\right] \, ,
\end{align}
where $ n_e(\mathbf{x})$ is the electron number density in units of 
$\si{cm^{-3}}$, $T_e(\mathbf{x})$ is the kinetic temperature of thermal electrons in $\si{K}$ and $\nu$ is the observing frequency in Hz. The optical depth $\tau$ in Eq.~\eqref{eq:flux_stimulated} is obtained by 
integrating the absorption coefficient along the line of sight from the Earth 
to the point where the stimulated axion occurs:
\begin{align}
    \label{eqn: optical depth}
    \tau(\nu,x,\Omega) = \int ^{x} _{0} \dd l\ \kappa(\nu,l,\Omega) \, .
\end{align}
Note that Ref.~\cite{Buen-Abad:2021qvj} used an alternative form for $\kappa$ by approximating the last term in brackets (Gaunt factor) in Eq.~\eqref{eqn:absorption coefficient} by a simple power law in $\nu$ and $T_e$. We use the exact expression~\eqref{eqn:absorption coefficient} for reasons explained in Section~\ref{subsec: electron profiles}.

So far we have implicitly assumed that both the DM and the
observer on Earth are at rest in the frame of the Galaxy. In reality there
are several factors that can affect the radio signal, especially when dealing
with Galactic point sources~\cite{Buen-Abad:2021qvj,Sun:2021oqp}:
\begin{itemize}
    \item DM in the Galaxy has a non-zero velocity dispersion 
	    $\sigma_{a} \approx 5 \times 10^{-4}c$~\cite{Freese:2012xd}, and
	a lower value in colder structures such as dwarf Spheroidal 
	satellites. As a result, axion decay does not occur at rest in the
	frame of the Galaxy and the photons do not point exactly along the
	line of sight from the Earth to a point source, but could deviate by an angle 
	$\theta$ up to $\sigma _{a}$ [cf.~Fig.~\ref{fig: stimulated decay}]. Hence, point sources are smeared 
	and the signal they generate would have a size 
	$2\psi \simeq 2\sigma _{a}(d _{s} + x)/d _{s}$, where $d _{s}$ is the 
	distance between the Earth and the source and $x$ is that between the 
	 Earth and the point where the axion decays. This effect is mostly relevant for point sources, and not for diffuse signals. Moreover, we will use the radio telescopes in the single-dish mode, and therefore, we can ignore the smearing effect, unlike Refs.~\cite{Buen-Abad:2021qvj, Sun:2021oqp} which use the interferometer mode where the angular size of the signal has to be carefully determined.  
 
    \begin{figure}[t!]
    \centering
    \includegraphics[width=0.99\textwidth]{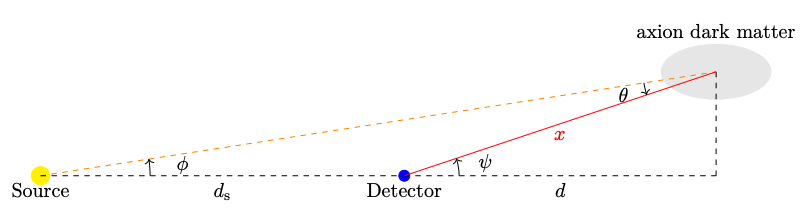}
    \caption{\label{fig: stimulated decay}A schematic diagram (not to scale) of the stimulated 
	   decay of an axion. The radio photon from the source (dotted orange 
	   line) induces the stimulated decay of an axion in a DM
	   cloud (the gray ellipse). Two back-to-back photons with energy equal to  
	   $m _{a}/2$ are emitted in the rest frame of the axion, and one of them (solid red line) is detected by a radio telescope on Earth (the pale blue dot), located at a distance $x$ from the point where axion decays. In the Galactic frame, the angle $\theta$ could 
	   be as large as the axion velocity dispersion $\sigma _{a}$. Using trigonometric identities, we can estimate the deviation angle $\psi \simeq \sigma _{a}(d _{s} + x)/d _{s}$.
    }
    \end{figure}
    The motion of axions also broadens the signal in frequency space, and
    we choose the observation bandwidth to maximize the signal-to-noise ratio 
    as discussed in Section~\ref{sec: telescopes}.

    \item The peculiar motion of a radio source causes different effects on 
	    the image depending on whether the motion is along or 
		perpendicular to our line of sight. 
		For instance, the signal could shift from the current location of the 
		source due to motion perpendicular to our line of sight. The 
		deviation can be
		estimated from the aberration angle $\theta _{\mathrm{de}} \simeq d _{p}/d _{s}$, 
		where 
		$d _{s}$ is the distance to the source and $d _{p}$ is the 
		distance 
		that the source travels during the relevant time to generate the signal. 
		
  For all the SNRs we consider here, $\theta_{\rm de}\lesssim 5$ arcmin,  which is typically smaller than the angular resolution of the radio telescopes like SKA in the single dish mode~\cite{Braun:2019gdo}, so the aberration effect will mostly cause the image of the SNR in the radio telescope to be blurred and enlarged by an order one factor.
		On the other hand, the motion of a source along 
		our line of sight can reduce or enhance the flux, which
		is proportional to the inverse of the distance between the 
		 Earth and the source squared. 
   This effect turns out to be 
		negligible for SNRs~\cite{Buen-Abad:2021qvj} and for the 
		Galactic center since the orbit of the solar system is 
		approximately circular, and therefore, the distance to the source remains nearly constant.

	\item The shadow of the Earth might temporarily shield the 
		radio photons from the source, which could either stop the stimulated
		decay of axions on the other side of Earth or prevent the photons from axion decays outside the shadow from passing the Earth. However, both effects turn out to be negligible due to the motion of the Earth in the DM rest frame~\cite{Buen-Abad:2021qvj}.

\end{itemize}

\section{The Galactic contribution}
\label{sec: the milky way}

We would like to determine the contribution to the radio signal 
generated by the stimulated decay of DM axions within the Galaxy in any direction. To evaluate the flux 
we first have to specify the distribution of the DM in the
Galaxy, the density and the temperature of the electron plasma that 
contributes to the opacity, and the background radio photons that
stimulate the decays of axions in the DM halo.

\subsection{The mass density of the Milky Way}
\label{subsec: mass density} 

The mass distribution of the Galaxy is typically described as the 
superposition of different components: the bulge, the thin and thick disks,
and the DM halo; see 
e.g. Refs.~\cite{mcmillan2011mass,mcmillan2016mass,binney2011galactic} for 
detailed models.

Numerical $N$-body simulations suggest that the density of the
DM in the Galaxy, which we assume to be exclusively 
composed of ALPs, 
follows the so-called NFW profile~\cite{Navarro:1996gj}:
\begin{align}
    \label{eqn:alp distribution}
	\rho_{\text{NFW}}(r)=\frac{\delta_{c} \rho_{c}}{\left(r / r_{s}\right)\left(1+r / r_{s}\right)^{2}} ,
\end{align}
where $r _{s}$ is the characteristic scale radius, $\rho _{c} \equiv 3H _{0}/8\pi G\simeq 10^{-29} {\rm g\ cm}^{-3}$ is the critical density of the Universe ($G$ being Newton's constant and $H_0$ the current Hubble parameter), and $\delta _{c}$ is  the overdensity parameter:
\begin{align}
    \delta_{\mathrm{c}}=\frac{\Delta_{\mathrm{vir}}}{3} \frac{C^{3}}{\ln \left(1+C\right)-C /\left(1+C\right)}, 
\end{align}
where the concentration parameter $C=R_{\rm vir}/r_s$,   and the halo mass is normalized by the virialization radius $R_{\rm vir}$ that encloses
a mass density $\Delta _{vir}=200 \rho_c$.  A recent comparison of the dynamics of the
Milky Way satellites with the predictions of Galaxy formation 
simulations~\cite{Callingham:2018vcf} yields a total mass of the halo of the Galaxy $M_{200}^{\rm MW}=
1.17^{+0.21}_{-0.15} \times 10^{12} M_\odot$ (at 68\% C.L.), and a concentration parameter of $C_{200}^{\rm MW} = 10.9^{+2.6}_{-2.0}$, from which we can derive $R_{\rm vir}\simeq 221$ kpc and $r_s \simeq 20$ kpc. From Eq.~\eqref{eqn:alp distribution}, this gives a local DM density of $\rho_{\rm local}\simeq 0.3~{\rm GeV \ cm}^{-3}$.

However, several observations suggest that while the relatively massive Galaxies show a cuspy NFW-like (or generalized NFW) profile~\cite{Cooke:2022upv}, a  cored DM 
profile provides a better fit to the rotation curves of smaller
galaxies~\cite{Flores:1994gz,Moore:1994yx,Moore:1999gc}. For example, the 
best NFW fit to the observations of dwarf Spheroidal satellites of the Milky
Way can be much less concentrated than expected from 
simulations~\cite{Maccio:2008pcd}, and mass estimates at different radii
are consistent with cored potentials~\cite{Walker:2011zu,Amorisco:2012rd}. 
Furthermore, baryonic physics processes such as stellar feedback are expected
to alter the central DM distribution in larger Galaxies like the
Milky Way~\cite{Chan:2015tna}. Thus, in our calculations we also consider 
the possibility that the DM is described by a cored Burkert 
profile~\cite{Burkert:1995yz}:
\begin{align}
	\rho_{\text{Bur}}(r)=\frac{\rho_s}{\pqty{1+\frac{r}{r_s}}
	\pqty{1+\frac{r^2}{r_s^2}}}.
\end{align}
Agreement with mass estimates of the Galaxy and requiring 
that the local DM density is
$\rho_{\rm local} \simeq 0.3~{\rm GeV \ cm}^3$ determines the scale radius 
$r _{\mathrm{s}} \simeq 12.67 \si{\ kpc}$ and the scale density 
$\rho_s \simeq 0.712~{\rm GeV \ cm}^3$~\cite{Cirelli:2010xx}.

\subsection{Electron number density in the Galaxy}
\label{subsec: electron profiles}
As shown in Eq.~(\ref{eqn:absorption coefficient}), the absorption 
coefficient depends on the number density and temperature of electrons along
the line of sight. Twelve models of the free electron distribution in the
Milky Way are explored in Ref.~\cite{Schnitzeler:2012jq}, and their parameters
are determined so that the observed distances of known pulsars are recovered
from their observed dispersion measure. Most models can predict the dispersion
measure within a factor of 1.5-2 for 75\% of the lines of sight. 
For definiteness, we use
the plane-parallel two-component model~\cite{Gomez:2001te}\footnote{Choosing
the TC93 model~\cite{1993ApJ...411..674T}, which provides a better fit to the pulsar data, has a dependence on the Galactic coordinates, and significantly
increases the computational costs, without altering the results much.}:
\begin{align}
    \label{eqn: electron number density}
n_e(R, z)=n_0 \frac{f\left(R / R_0\right)}{f\left(R_{\odot} / R_0\right)} f\left(\frac{z}{z_0}\right)+n_1 \frac{f\left(R / R_1\right)}{f\left(R_{\odot} / R_1\right)} f\left(\frac{z}{z_1}\right)
\end{align}
where $R$ is the polar distance from the Galactic Center, $R_{\odot}=8.5~{\rm kpc}$ is the Galactocentric distance of the Sun, and $z$ is the height 
above the Galactic mid-plane. We take the 
best-fit parameters from Table 4 in Ref.~\cite{Schnitzeler:2012jq} for $f(x) = $ sech$^2(x)$, i.e.,
\begin{align}
   & n_0 = 1.77\times 10^{-2}~{\rm cm}^{-3}, \quad 
    R_0 = 15.4~{\rm kpc}, \quad z_0= 1.10~{\rm kpc}, \\
   & n_1 = 1.07\times 10^{-2}~{\rm cm}^{-3}, \quad 
   R_1= 3.6~{\rm kpc}, \quad z_1=0.04~{\rm kpc}.
\end{align}
We note that our calculation of the optical depth using the electron number density given above is different from Ref.~\cite{Buen-Abad:2021qvj}, which used the emission measure, ${\rm EM} = \int n_e^2 \dd x$, along the line of sight, with $n_e$ modeled as a plane-parallel distribution with a single scale height~\cite{Berkhuijsen:2008zv, Gaensler:2008ec}.  This approach makes the optical depth negligible for the entire range of frequencies they investigated. In our case, the optical depth is negligible only at higher frequencies ($\gtrsim 500$ MHz), but is important at low frequencies, where we get the best sensitivities. Moreover, the single-plane model does not provide a good fit to the pulsar data~\cite{Schnitzeler:2012jq}. Both models greatly underestimate the EM in the close vicinity of the Galactic Center, which has been estimated to be $\sim 10^5~{\rm cm}^{-6}{\rm pc}$~\cite{1989ApJ...342..769P}, based on the radio observation of Sgr A$^*$ and the 7 arcmin halo surrounding it. But the two component model gives ${\cal O}(10^2)~{\rm cm}^{-6}{\rm pc}$, compared to ${\cal O}(10^{-1})~{\rm cm}^{-6}{\rm pc}$ in the single-component model. See Section~\ref{sec:skymaps} on how we handle the EM discrepancy at the Galactic Center emission.

As for the electron temperature in Eq.~\eqref{eqn:absorption coefficient}, it can be derived from observations of radio 
recombination line and continuum emissions in H II regions. Taking a sample of 76 nebulae widely distributed over the Galactic disk, the best fit electron temperature in the Galaxy was found to be  $T_e=(5780\pm 350)+(287\pm 46)(R_{\rm gal}/{1~{\rm kpc}})$ K, where $R_{\rm gal}$ is the Galactocentric distance~\cite{Quireza:2006sn}. Following Ref.~\cite{Caputo:2018vmy}, we will ignore the small temperature gradient in the Galactic disk and will assume a constant electron temperature $T _{e} \sim 5000 \si{\ K}$.

\subsection{Diffuse radio emission}
\label{subsec: radio emission}
As discussed in Section~\ref{sec: stimulated decay of axion}, 
the presence of a background of radio photons 
will stimulate the decay of DM axions, and the resulting flux of 
photons is linearly proportional to the occupation number $f_\gamma$ of 
background photons. When estimating the diffuse emission from an arbitrary
direction, we consider three separate contributions to $f_\gamma$: the 
cosmic microwave background (CMB), the extragalactic radio background (ERB), 
and the Galactic radio emission.

At higher frequencies ($\gtrsim 10 \si{\ GHz}$), the main contribution to
the photon background comes from the CMB. The spectrum is, hence,
well approximated by a black-body spectrum:
\begin{align}
    f _{\gamma,\mathrm{CMB}}(E_\gamma) = \frac{1}{e ^{E _{\gamma}/k_B T_\mathrm{CMB}} - 1} ,
\end{align}
with $T _{\mathrm{CMB}} = 2.72548 \pm 0.00057  \si{\ K}$~\cite{Fixsen:2009ug}. Here $k_B$ is the Boltzmann constant. 

Since extragalactic radio sources are likely to be distributed isotropically, 
the observed ERB is expected to be nearly isotropic. 
The analysis of radio maps with frequencies ranging from $22 \si{\ MHz}$ to $2.3 \si{\ GHz}$ provides a best fit to the brightness temperature of 
the isotropic extragalactic radio emission~\cite{Fornengo:2014mna}:
\begin{align}
    T_{\text {exgal}}(\nu) \simeq 1.19~{\rm K}\left(\frac{1~\mathrm{GHz}}{\nu}\right)^{2.62} .
    \label{eq:ERB}
\end{align}
Then the stimulated enhancement factor can be obtained from the measured radio intensity:
\begin{align}
    f_{\gamma}(E_\gamma) = \frac{\pi^2\rho_\gamma}{E_\gamma^3} \, ,
\end{align}
where the energy density (which is same as the intensity up to a factor of $c$, the speed of light) can be written in terms of the brightness temperature as\footnote{Here we keep the Planck's constant $h$ for clarity, although elsewhere in the text, we have assumed $\hbar\equiv h/2\pi=1$.}  
\begin{align}
    \rho_\gamma= \frac{2h\nu^3}{e^{h\nu/k_B T_b}-1} \simeq 2 \nu^2 k_B T_b\quad \text{(for $h\nu \ll k_BT$)} \, .
\end{align}
For the frequency range of our interest, $h\nu \ll k_BT_b$ holds for the extragalactic brightness temperature given in Eq.~\eqref{eq:ERB}, because $h\nu \sim m_a \sim \mu\text{eV}$, whereas $k_BT \sim 1~{\rm K}\sim 10^{-4}~{\rm eV}$.

From the above discussion, we see that both CMB and ERB sources contribute an isotropic radio background to the stimulated photon signal which only depends on the photon energy, or equivalently, on the axion mass. But the Galactic contribution to the radio background is more complicated, and depends on the Galactic coordinates of the line of sight. Radio photons are produced in the Galaxy via several processes including 
synchrotron radiation, thermal bremsstrahlung, and synchrotron self-absorption
in SNRs. We estimate the Galactic contribution to $f_\gamma$
from the observed brightness temperature $T_{\mathrm{gal}}$ 
of the $408 \si{\ MHz}$ Haslam map~\cite{haslam1981408,haslam1982408}. 
Although the spectral index of the Galactic radio emission generally depends 
on the direction~\cite{guzman2011all}, we will use the ansatz given in Table 1 of Ref.~\cite{Yusef-Zadeh:2012efm} that was also used in Ref.~\cite{Caputo:2018vmy}:
\begin{align}
    f_{\gamma,\text{gal}}(r, \nu)=\left.f_{\gamma, \mathrm{gal}}(r)\right|_{\nu=\nu_*} \times\left\{\begin{array}{ll}
    \left(\nu / \nu_*\right)^{-3.173} & \nu<\nu_* \\
    \left(\nu / \nu_*\right)^{-3.582} & \nu_* \leq \nu \leq 4.85 \ \mathrm{GHz} \\
    1.99 \left(\nu / \nu_*\right)^{-4.14} & \nu>4.85 \ \mathrm{GHz}
    \end{array}\right. \, ,
    \label{eq:fgal}
\end{align}
which is obtained by fitting the observed radio flux from the inner $2 ^{\circ} \times 1 ^{\circ}$ of the Galactic Center region at $\nu _{\star} = 1.415 \si{\ GHz}$.\footnote{Our Eq.~\eqref{eq:fgal} differs from Eq.~(4.7) in Ref.~\cite{Caputo:2018vmy} for $\nu>4.85$ GHz and ensures that $f_{\gamma,{\rm gal}}$ is continuous at $\nu=4.85$ GHz. } As we will see in Section~\ref{sec: results}, the gegenschein signal induced by the diffuse emission from the Galactic Center (or Anti-center for low frequencies) give us better sensitivity than the point sources.

\subsection{Radio emission from SNRs}
\label{subsec: radio emission from SNRs}

Our objective is to determine the most favorable location in the sky to 
measure a potential radio signal associated with the stimulated
decay of axion DM. Besides the diffuse emission (both direct and gegenschein) from the 
central region of the Galaxy, the gegenschein signal associated with bright 
point sources also stand out. SNRs are of particular interest, because they are known to be 
radio bright~\cite{vink2012supernova, Dubner:2015aqa, vink2020physics}. SNRs could emit a copious number of radio photons both thermally and non-thermally. Thermal radio emission consists of bremsstrahlung (free-free emission), photon emissions from charged particles accelerated by encountering another charged particle, radiative recombination continuum (free-bound emission),  single photon emission due to the capture of a free electron by an ion, and two-photon emission from electrons in metastable states. Nonthermal emission includes synchrotron radiation caused by relativistic charged particles, mainly electrons and positrons, gyrating in a magnetic field and scattering between photons and cold electrons (Thomson scattering). Emissions due to synchrotron radiation are dominant in the radio frequencies of our interest and the photon spectrum follows a simple power law: $S_\nu\propto E_\gamma^{-\alpha}$, where the spectral index $\alpha$ is typically around 0.5; see Table~\ref{tab: snr parameters}.

As implied by Eq. (\ref{eq:flux_stimulated}), a source that was brighter in the past could 
produce a brighter signal compared to its present state. As first proposed 
in Ref.~\cite{Woltjer:1972at}, and summarized in~\cite{draine2010physics,truelove1999evolution}, the time evolution of SNRs  can be roughly divided in four phases, namely, (i) Free expansion (or ejecta-dominated) phase, (ii) Adiabatic expansion (or Sedov-Taylor) phase, (iii) Radiative (or Snow-plough) phase, and (iv) Dispersion (or merging) phase. After a short
initial free expansion phase that lasts a few hundred years, the bulk of the radio emission is generated 
in the early stages of the Sedov-Taylor or adiabatic phase. During this 
phase, which lasts $\sim {\cal O}(\textrm{few}\times 10^4$) years, the luminosity decreases 
steeply with time, so the total integrated gegenschein luminosity is expected
to be much greater than what we would infer from the source's luminosity at
present, since most of the radio emission was produced when the source was
young.

The important SNR evolution parameters are the spectral index $\alpha$, magnetic field amplification (MFA) time $t_{\rm MFA}$, and the age $t_0$. Following Ref.~\cite{Sun:2021oqp}, we will consider two models for the evolution of the spectral index in the Sedov-Taylor phase, namely, 
(i) $S_\nu\propto t^{-4\gamma/5}$~\cite{shklovskii1960secular}, and (ii) $S_\nu\propto t^{-2(\gamma+1)/5}$~\cite{Urosevic:2018uoc}, where $\gamma$ is the power law index for the differential energy spectrum of the synchrotron electrons: $\dd n_e/\dd E_e\propto E_e^{-\gamma}$, and is related to the spectral index of the photon flux by $\gamma=2\alpha+1$~\cite{rohlfs2013tools}. Thus, $\gamma>1$ (since $\alpha>0$), which implies the model (i) gives more optimistic flux. Similarly, for $t_{\rm MFA}$, we will use the range of 30--300 years, with a central value of 100 years~\cite{2009ApJ...695..825I, Pavlovic:2017asa}.

Out of the roughly 300 Galactic SNRs recorded in Green's 
catalog~\cite{Green:2019mta,Green:url}, 60 of them have known distance, age
and spectral index listed in the SNRcat 
catalog~\cite{Ferrand:2012jh,Ferrand:url}. Among these, the counterimage
of W50 was found in Ref.~\cite{Buen-Abad:2021qvj} to yield the maximum 
signal-to-noise ratio in SKA 1, while Ref.~\cite{Sun:2021oqp} also considered
Vela and W28. In our survey we add S147 to this list (see Table~\ref{tab: snr parameters}), because S147 (G180.0-1.7) is the closest SNR to the Galactic Anti-center, and therefore, its counterimage will be formed close to the Galactic Center. Although this  
counterimage is likely to be contaminated by radio photons from the Galactic 
Center, we still expect a large gegenschien flux from it 
since the flux from axion decay is proportional to the density of axions, which is the largest at the Galactic Center. 
Furthermore, S147 is the oldest of the four SNRs considered here, which enables its radio photons to stimulate decays of 
axions located far away from the Earth. Its spectral index varies depending on frequency: below $1.7 \si{\ GHz}$, the spectrum is relatively flat, and the spectral index $\alpha$ is estimated to be $0.3$, while it grows to $1.2$ above
$1.7 \si{\ GHz}$. 

W28 (G6.4-0.1) is almost on the Galactic plane, but its counterimage would 
avoid being smeared by the radio background since it is away from the direction of the Galactic Center. It is encouraging that this SNR is almost as old as S147, 
and its observed flux at $1 \si{\ GHz}$ is large. On the other hand, Vela (G263.9-3.3) is a relatively young SNR, but it is the brightest among the four, even after excluding the radio photons from the pulsar nebula, and has a large spectral index, and thus, is expected to give a bright counterimage. 
Finally, W50 (G39.7-2.0) is the SNR with which Refs.~\cite{Sun:2021oqp,Buen-Abad:2021qvj} place the strongest constraints on the axion-photon coupling $g _{a\gamma}$.  It has a large spectral index, a
relatively long age and is furthest away (among the four), which is a good combination that helps to generate a bright gegenschein image. 
As we will see in Section~\ref{sec: results}, S147 gives a slightly better constraint than the other three point sources considered before, but the diffuse emission from the Galactic Center and Anti-center give us the best constraints.
\begin{table}[t!]
    \centering
    \begin{tabular}{ccccc} \hline
        & S147 & W28 & Vela & W50 \\  \hline \\ [-10pt]
         Distance [pc] & 1470$\substack{+420\\-270}$ & 1900$\substack{+300\\-300}$ & 287$\substack{+19\\-17}$ & 50000$\substack{+5000\\-5000}$ \\  [2pt]
        Age [kyr] & 40$\substack{+20\\-10}$ & 34.5$\substack{+1.5\\-1.5}$ & 12$\substack{+2\\-2}$ & 
        30$\substack{+70\\-10}$
        \\[2pt]
        Spectral index $\alpha$ & 0.3$\substack{+0.15\\-0.15}$ (1.2$\substack{+0.3\\-0.3}$) & 0.42$\substack{+0.02\\-0.02}$ & 0.74$\substack{+0.04\\-0.04}$ & 0.7$\substack{+0.1\\-0.2}$ \\ [2pt]
        Flux $S _{\nu}$ [Jy] & $59\pm 6$ & $310\pm 20$ & $610\pm 105$ & $85\pm 20$ \\ [2pt]
        Galactic coordinates $(l,b)$ & $(180.0, -1.7)$ & $(6.4, -0.1)$ & $(263.9, -3.3)$ & $(39.7, -2.0)$ \\ [2pt]   
        Size [arcmin] & 180 & 48 & 255 & 60 \\ \hline
    \end{tabular}
    \caption{\label{tab: snr parameters}Parameters of the four SNRs used in 
	this paper and the associated uncertainties~\cite{Green:2019mta,Green:url} (see also Ref.~\cite{Sun:2021oqp} and references therein). The spectral index of 
	S147 depends on the frequency; the central value is 0.3 (1.2) for frequencies 
	smaller (larger) than $1.7 \si{\ GHz}$~\cite{reich200335}. The fluxes are given at 
	$\nu = 1.7 \si{\ GHz}$ for S147, and at $\nu = 1 \si{\ GHz}$ for 
	the others. The flux uncertainties quoted here are based on the available measurements closest to 1 GHz. 
}
\end{table}

The SNR parameter values and uncertainties for the four point sources used in this study are taken from the Green's catalog~\cite{Green:url} and are summarized in Table~\ref{tab: snr parameters}.   

\subsection{Effects of small scale structure}
\label{subsec: gravitational interactions of axions}

Axion models where the PQ symmetry is broken after inflation 
generically predict that the DM distribution will be inhomogeneous
at small scales. As a result, a fraction of the DM mass collapses
into small halos called axion 
miniclusters~\cite{Hogan:1988mp,Kolb:1993zz,Kolb:1993hw,Vaquero:2018tib,Buschmann:2019icd,Eggemeier:2019khm,Ellis:2020gtq,Xiao:2021nkb, Ellis:2022grh, Dandoy:2022prp, Pierobon:2023ozb}. Even denser axion
stars~\cite{Kaup:1968zz,Ruffini:1969qy,Seidel:1991zh,Visinelli:2021uve} might 
form within miniclusters in a time shorter than the age of the 
Universe~\cite{Tkachev:1991ka,Levkov:2018kau,Eggemeier:2019jsu,Dmitriev:2023ipv}. The presence of these inhomogeneities can give rise to striking 
effects~\cite{Levkov:2020txo,Buckley:2020fmh,Edwards:2020afl,Visinelli:2021uve,Amin:2021tnq,Bai:2021nrs,Iwazaki:2022bur,Witte:2022cjj,Fox:2023aat,Escudero:2023vgv,Dandoy:2023zbi} and it is worth exploring whether their presence
can also be revealed by their stimulated radio decays.

Since the precise abundance and distribution of axion stars is uncertain,
we choose for our purposes a particular realization of the DM of
the Galaxy where a fraction of order 10\% of the DM mass is in the
form of dilute axion stars with mass $m_{\rm AS}$ and radius~\cite{Chavanis:2011zm}
\begin{equation}
        R_{a}^{\text {dilute }}  \sim  27000 \mathrm{~km} \;
	\pqty{\frac{1~ \mu \mathrm{eV}}{m_{a}}}^{2}
	\pqty{\frac{10^{-12} M_{\odot}}{m_{\rm AS}}}.
	\label{eq:dilute}
\end{equation}
In principle, the presence of such a clumpy component could give rise to
statistical fluctuations of the radio signal along different lines of sight.
However, the solid angle of the telescope beam corresponds to
a DM column mass that is several orders of magnitude larger than the 
mass of the individual axion clumps~\cite{Sun:2021oqp}. For
instance, for typical dilute star with radius given in Eq.~(\ref{eq:dilute})
we would expect about $10^{17}$ dilute stars along a given line of sight,
so the tiny $\sim 10^{-8}$ Poisson fluctuations are unobservable.

Moreover, since the signal from stimulated axion decay is proportional to
the DM density, the presence of a clumpy component 
does not appear to change the observed flux (unlike e.g. an indirect signal 
from DM annihilation, which is proportional to the square of the 
density). This is indeed the case as long as the clumpy DM component
has the same density distribution as the underlying homogeneous DM
profile. However, if clumps are preferentially located e.g. at the outer
parts of the halo compared to the homogeneous NFW or Burkert profile, the
flux from stimulated axion decay would be different than the one expected
from a completely smooth DM halo.

Since the spatial extent of axion miniclusters is small,
$\lesssim 1$ pc, and that of dilute stars is even smaller [cf.~Eq.~\eqref{eq:dilute}],  their initial mass distribution at the time the DM 
halo of the Galaxy was assembled is likely to coincide with that of the
homogeneous halo.
However, their distribution might be modified over time due to the 
effects of dynamical friction between ordinary stars and axion stars. 
Gravitational interactions between collisionless components lead to 
equipartition, where the average kinetic energy of the lighter component 
becomes equal to that of the heavier 
component~\cite{Spitzer:1940xd,Chandrasekhar:1943ys,Spitzer:1969xd}. 
This leads to mass segregation, where the more massive component (i.e. the
ordinary star in this case) tends to drift closer to the cluster. 
Assuming that both axionic and ordinary stars follow a Maxwellian velocity 
distribution, the mean transferred energy is given by~\cite{Merritt:2013xd}
\begin{align}
    \label{eqn: energy transfer}
    \frac{\dd  E_{\rm AS}}{\dd  t}=\frac{\sqrt{96 \pi} G^{2} m_{\rm AS} \rho_{\mathrm{s}} \ln \Lambda}{\left[\left\langle v_{\rm s}^{2}\right\rangle+\left\langle v_{\mathrm{AS}}^{2}\right\rangle\right]^{3 / 2}}\left[m_{\mathrm{s}}\left\langle v_{\mathrm{\rm s}}^{2}\right\rangle-m_{\rm AS}\left\langle v_{\rm AS}^{2}\right\rangle\right],
\end{align}
where subscripts denote either stars (s) or axion stars (AS), $\langle v_{\rm s}^2\rangle$ and $\langle v_{\rm AS}^2\rangle$ are the corresponding mean-squared velocities, and $\rho _{\rm s} = \rho _{\rm b} + \rho _{\rm d,\mathrm{thin}} + \rho _{\rm d,\mathrm{thick}}$ is
the density of ordinary stars which receives contributions from the bulge and from the thin and thick discs~\cite{mcmillan2011mass,  mcmillan2016mass}. 
We assume that the axion stars have the same velocity dispersion as that of the homogeneous component
of the DM~\cite{Freese:2012xd}, 
which is comparable to velocity dispersion of stars in the 
Galaxy~\cite{Battaglia:2005rj, Brown:2009nh}, i.e. $\langle v_{\rm s}^2\rangle\sim \langle v_{\rm AS}^2\rangle = \sigma_a^2 \simeq (5 \times 10^{-4}c)^2$.  
The Coulomb logarithm in Eq.~\eqref{eqn: energy transfer} is approximated by $\ln \Lambda=\ln[d_{\max } \sigma_a^{2}/(G\left(m_{\mathrm{s}}+m_{\mathrm{AS}}\right))]$, where $d _{\mathrm{max}}$ is an upper limit to the distance between a 
gravitationally interacting axion star and an ordinary star. A reasonable choice for 
$d _{\mathrm{max}}$ would be the bulge radius, $r_b\simeq  2.1 \si{\ kpc}$. This gives $\ln\Lambda = 23.1$.  We can now estimate the relaxation time 
$t _{\mathrm{relax}}$ as 
\begin{align*}
    t _{\mathrm{relax}} &\equiv\left|\frac{1}{\frac{1}{2}m_{\mathrm{AS}} \left\langle v_{\mathrm{AS}}^{2}\right\rangle} \frac{\mathrm{d}}{\mathrm{d} t}\left(\frac{1}{2}m_{\mathrm{AS}} \left\langle v_{\mathrm{AS}}^{2}\right\rangle\right)\right|^{-1} \simeq \frac{0.08 \  \sigma_a^{3}}{G^{2} m_{\mathrm{s}} \rho_{\mathrm{s}} \ln \Lambda} \atag \simeq 7000 ~ \mathrm{Gyr} \left(\frac{100 \ M_{\odot} \mathrm{pc}^{-3}}{\rho_{\text {s}}}\right)\left(\frac{M_{\odot}}{m_{\rm s}}\right),
    \label{eq:trelax}
\end{align*}
which turns out to be much longer than the age of the Universe for typical values of the stellar density in the bulge, $\rho_s\simeq 100 \ M_{\odot} \mathrm{pc}^{-3}$, and a typical stellar mass of $m_{\rm s}\simeq 1\  M_{\odot}$.  
Hence, the mass segregation of axion stars and ordinary stars due to dynamical friction is an
extremely slow process for the given mass hierarchy between ordinary and
axionic stars. 

In order to estimate the current distribution of axion stars, we further 
assume that the effect of mass segregation is spherically symmetric so that
we can apply the virial theorem, $E = -U/2$ (where $E$ and $U$ are the kinetic and potential energies respectively), to the system. Under these 
assumptions, the evolution of a radial mass shell is governed by the 
following differential equation~\cite{Brandt:2016aco}:
\begin{align}
    \label{eqn: radial mass shell evolution}
    \frac{\dd r}{\dd t}=\frac{4 \sqrt{2} \pi G ^{2} \rho _{{\rm s}} m _{{\rm s}}}{\sigma} \ln \Lambda \left( \dv{f(r)}{r}\right)^{-1}
\end{align}
where $f(r)$ is the gravitational potential energy per unit mass as a 
function of distance $r$ from the Galactic Center. Here the gravitational potential is taken to be the sum of the 
contributions from the bulge, the thin disk, the thick disk, the axion stars, 
and the supermassive black hole at the Galactic Center with mass 
$M_{\rm BH} \approx 4 \times 10^6 M_\odot$ \cite{abuter2019geometric}. 

\begin{figure}
    \centering
    \includegraphics[width=0.5\textwidth]{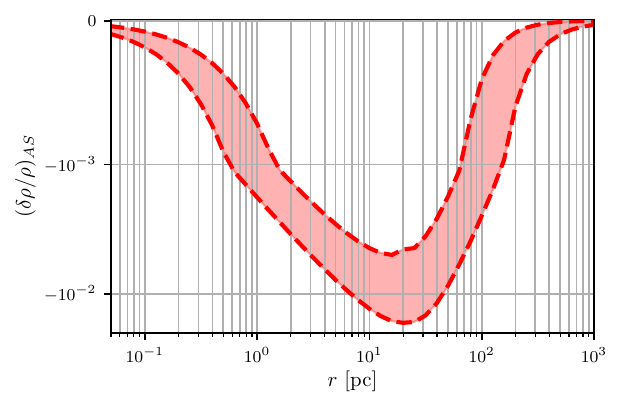}
    \caption{\label{fig: massdeficit} Mass deficit due to mass segregation between axion stars and ordinary stars as a function of distance from the Galactic Center. The band shows the uncertainty in the stellar mass density distribution in the Galaxy. }
\end{figure}

Figure~\ref{fig: massdeficit} shows the mass density deficit 
$(\delta \rho/\rho)_{\rm AS} \equiv (\rho_{\rm AS}(t) - \rho_{\rm AS}(0))/\rho_{\rm AS}(0)$ for the axion stars, calculated by solving Eq.~(\ref{eqn: radial mass shell evolution}), and taking $\rho(t)/\rho(0) \simeq (r(0)/r(t))^3$ at $t=13.8$ Gyr, the current age of the Universe. 
We use two models to compute the mass deficit: For the optimistic model (lower curve) we set the stellar mass density $\rho_{\rm s}(r)$ to the
value on the Galactic plane, whereas for the conservative model (upper curve) we take it equal to the value in the direction of the Galactic North (or South) pole. 
We see that the mass density of axion stars is reduced by at most 0.5\% (conservative)-1.7\% (optimistic) at 
$r \sim 20 \si{\ pc}$ from the Galactic Center. This will causes a reduction in the photon flux from stimulated axion decays by about 0.1\% (conservative)-0.3\% (optimistic). Thus, the mass segregation effect on the radio signal is negligible.

\section{Radio telescopes}
\label{sec: telescopes}

As shown in Eq.~\eqref{eq:radio}, the frequency of photons produced from stimulated axion decay naturally falls in the radio band for $m_a\sim \mu$eV, where the axion can make up the bulk 
of the DM. There are several radio telescopes that are currently operating, and even 
larger facilities will start collecting data in the next few years~\cite{2013pss1.book..315E}. We here
 summarize the characteristics of FAST~\cite{Nan:2011um} (currently operating) and SKA~\cite{Braun:2019gdo} (Phase 1 under construction and Phase 2 planned) as representative examples, which will be used to derive the sensitivity curves for the axion masses and couplings from the stimulated decay signal.

The Square Kilometer Array (SKA)~\cite{5136190} has one of the largest collecting areas among
all radio telescopes. It comprises two different types of instruments, SKA-low
and SKA-mid, that use dipole and parabolic antennas, respectively. 
SKA-low, located in Western Australia, observes radio photons with lower frequencies (50-350 MHz)  
 and it consists of $512$ telescopes, each with $35 \si{\ m}$ diameter and $256$ antennas.
SKA-mid, located in South Africa, is sensitive to photons 
with higher frequencies (350 MHz - 1.54 GHz, which will extend to 50 GHz with SKA2). It consists of $133$ dishes with a diameter of 
$15 \si{\ m}$ and $64$ MeerKAT dishes with a diameter of 
$13.5 \si{\ m}$ \cite{braun2014ska1,braun2015ska1}. 
On the other hand, FAST~\cite{Nan:2011um} is a single-dish telescope with a diameter of 300 m currently operating in China. Its design frequency range is from 70 MHz to 3 GHz, and up to 8 GHz with future upgrades. We summarize the specifications of these telescopes in 
Table~\ref{tab: parameters of telescopes}.

\begin{table}[t!]
\centering
\begin{tabular}{cccccc} \hline
    & SKA1-low & SKA2-low & SKA1-mid & SKA2-mid & FAST\\ \hline
    Frequency [MHz] & 50-350 & 50-350 & 350-15400 & 350-50000 & 70-3000\\
    $N _{\mathrm{tele}}$ & 512 & 4800 & 197 & 2000 & 1\\
    $D$ [m] & 35 & 35 & 13.5, 15 & 13.5, 15 & 300\\ 
    $\theta _{\mathrm{res}}$ [deg] & 12-1.7 & 12-1.7 & 4.0-0.91 & 4.0-0.28& 1.0-0.023\\ 
    $T _{\mathrm{r}}$ [K] & 40 & 40 & 20 & 20 & 20\\ \hline
\end{tabular}
\caption{\label{tab: parameters of telescopes} Main properties of the
	SKA~\cite{ska} and FAST~\cite{Jiang:2019rnj} relevant to our calculation.}
\end{table}

Several key properties of a telescope enter the calculation of the axion decay
signal: angular resolution, noise properties, bandwidth and frequency 
resolution. The angular resolution of a telescope with a diameter $D$ is given by
\begin{align}
    \theta _{\mathrm{res}} \simeq 1.22 \frac{\lambda}{D} \simeq 1.4^{\circ}\left(\frac{1~\mathrm{GHz}}{\nu}\right)\left(\frac{15 \mathrm{~m}}{D}\right).
\end{align}
Given the resolution $\theta _{\mathrm{res}}$, we can express the primary beam
angular size as $\Omega_{\rm pb}=2 \pi\left(1-\cos \left(\theta _{\mathrm{res}} / 2\right)\right)$. 

The signal power in a bandwidth $\Delta \nu$ observed by each antenna can be expressed as
\begin{align}
    P_{\text {signal }}&=\eta A f_{\Delta} S_{\nu} \Delta \nu ,
\end{align}
where $S_\nu$ is the observed flux of radio photons in the bandwidth $\Delta \nu$ [cf.~Eq.~\eqref{eq:flux_stimulated}] with the integral over solid angle extending over the primary beam area, $A=\pi(D/2)^2$ is the area of each dish, 
$\eta$ is the detector efficiency which we set conservatively to 
$0.8$~\cite{Braun:2019gdo} for SKA and $0.7$ for FAST~\cite{Nan:2011um}.

Assuming that axion DM 
follows a Maxwell-Boltzmann velocity distribution, 
the signal-to-noise ratio is maximized for
$\Delta \nu = 2.17 \nu _{a} \sigma _{a}$,  which contains 
a fraction $f _{\Delta} = 0.721$ of all the photons from stimulated axion
decays~\cite{Ghosh:2020hgd}. Our signal bandwidth $\Delta \nu/\nu_a = 2.17 \sigma_a \sim 10^{-3}$ is within the observable range of the radio telescopes considered here.  

The instrument noise is characterized by the power 
\begin{align}
    P_{\text {noise}}=2 k_{B} T \sqrt{\frac{\Delta \nu}{t_{\mathrm{obs}}}},
\end{align}
where $t _{\rm obs}$ is the observation time that we take to be 100 hours for definiteness.  
$T = T_{\mathrm{a}}+T_{\mathrm{CMB}}+T_{\mathrm{exgal}}+T_{\mathrm{gal}}+T_{\mathrm{r}}$ is the noise temperature which is the sum of contributions from
atmospheric radio photons, CMB, extragalactic 
radio waves, Galactic radio emission, and the temperature of the receiver. 
The brightness temperature of the atmospheric signal is set to 
$T _{\mathrm{a}} = 3 \si{\ K}$~\cite{ajello1995evaluation}. The receiver temperature is $T _{\mathrm{r}} = 20 \si{\ K}$ for FAST~\cite{Nan:2011um}  and SKA-Mid, and $40 \si{\ K}$ for SKA-Low~\cite{Braun:2019gdo}. As discussed in Section~\ref{subsec: radio emission} we 
take the frequency-dependent $T_{\text {exgal}}$ from Eq.~\eqref{eq:ERB}, and that of the Galactic emission is taken from the Haslam map~\cite{haslam1981408, haslam1982408}.

Putting things together, we find that
the signal-to-noise ratio for a single telescope is 
given by  
\begin{align}
    \label{eqn: signal-to-noise ratio for a telescope}
    \qty(\frac{S}{N}) _{\mathrm{single}} =\frac{m_a^3 g _{a\gamma} ^{2}}{512\pi^2} \frac{\eta A f_{\Delta}}{k _{B}T}\sqrt{\frac{t_{\mathrm{obs}}}{\Delta \nu}} \sqrt{n_{\rm pol}} \int\dd x \int \dd \Omega\ \rho _{a}(x,\Omega) e ^{-\tau(\nu_a,x,\Omega)} \left(f_{\gamma}(x,\Omega,t)+\tilde{f}_{\gamma}(x, \Omega, t) \right) ,
\end{align}
where $n_{\rm pol}=2$ is the number of polarizations of photons. 
If we now consider an array of $N _{\mathrm{tele}}$ telescopes, 
the signal-to-noise ratio is the root mean square of the signal-to-noise ratio of each 
telescope. Therefore,
\begin{align}
    \label{eqn: signal-to-noise ratio for an array}
    \qty(\frac{S}{N}) _{\mathrm{array}} = \sqrt{N _{\mathrm{ant}}}\qty(\frac{S}{N}) _{\mathrm{single}}
\end{align}
where $N _{\mathrm{ant}}=N_{\rm tele}n_{\rm ant}$ is the total number of antennas, i.e. the product of the number of telescopes and the number of antennas in each telescope.

A pair of telescope dishes can also work as an interferometer with an angular resolution 
$\theta_{\rm pair} \simeq \lambda/d_{\rm pair}$ where $d_{\rm pair}$ is the spatial 
separation of the two dishes. In particular, the SKA telescope can be operated
as a radio interferometer with $N_{\rm tele}(N_{\rm tele}-1)/2$ pairs of dishes. Each two-element interferometer contributes to the 
measurement when the angular size of the source $\theta_{\rm source}$ is smaller 
than the angular resolution $\theta_{\rm pair}$. For larger sources, the 
visibility $R \propto \sin q/q$ where $q\equiv \pi \theta_{\rm source}/\theta_{\rm pair}$ becomes smaller and not all two-element interferometers 
contribute~\cite{rohlfs2013tools}. The maximum sensitivity is attained when 
all pairs of dishes function as interferometers. In general, the signal-to-noise 
ratio falls off like $S/N \propto \sqrt{N_{\rm useable}/N_{\rm pair}}$ for extended 
sources where $N_{\rm useable}$ is the number of useable pairs. 

For extended sources like the Galactic center, with angular size of  
$\sim \mathcal{O}(1^\circ)\times\mathcal{O}(1^\circ)$, single-dish mode is 
likely to achieve a larger signal-to-noise ratio. Even for SNRs, 
single dish mode gives better sensitivity \cite{Buen-Abad:2021qvj} at lower 
frequencies. Thus, we will only use the single-dish mode for SKA.

\section{Results}
\label{sec: results}

The non-observation of photons from the stimulated decays of axions in
the Galactic DM halo can be used to place constraints on the axion-photon coupling 
$g _{a\gamma}$. For definiteness, we assume 100 hours of observation time and set 
the signal-to-noise ratio threshold $(S/N)_{\rm array}>1$ to derive our sensitivity limits using FAST and SKA radio telescopes.

\begin{figure}
    \centering
    \vspace{-30pt}
    
    \includegraphics[width=0.8\textwidth]{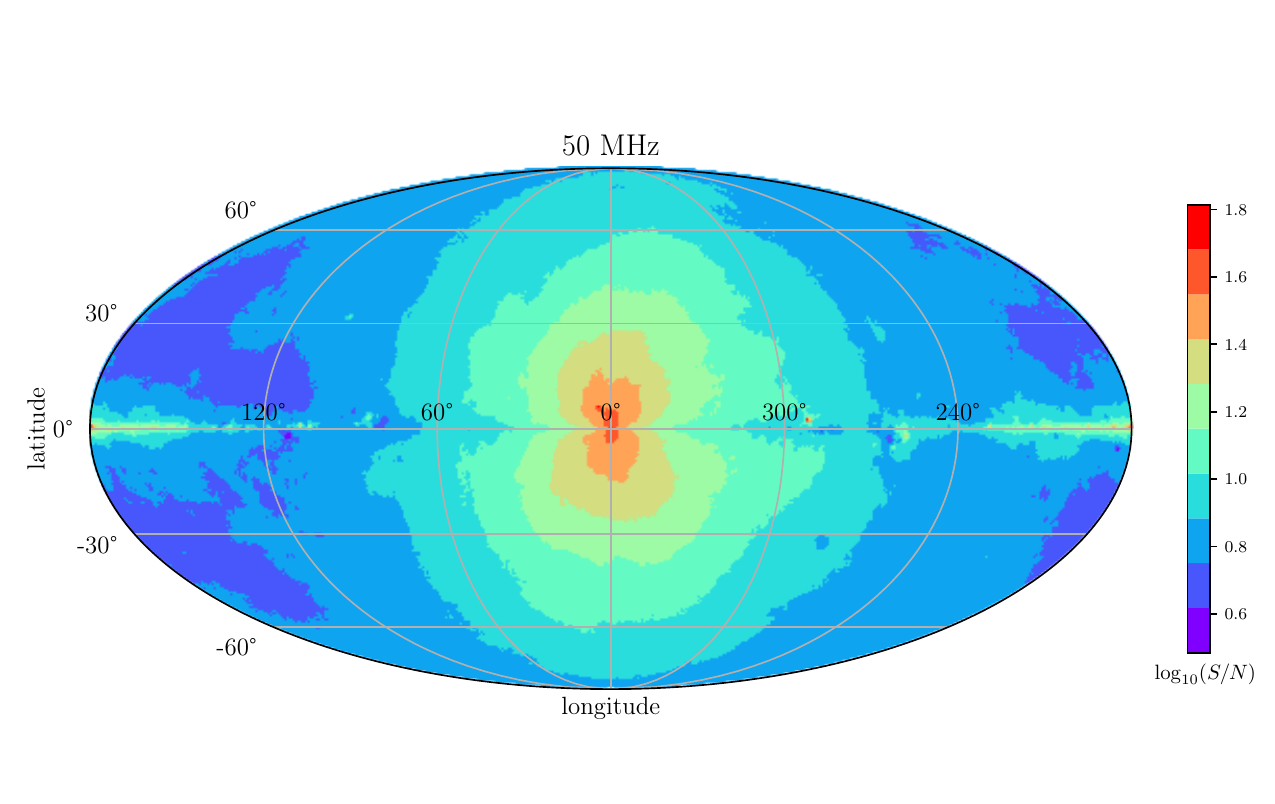}
    \vspace{-60pt}
    
    \includegraphics[width=0.8\textwidth]{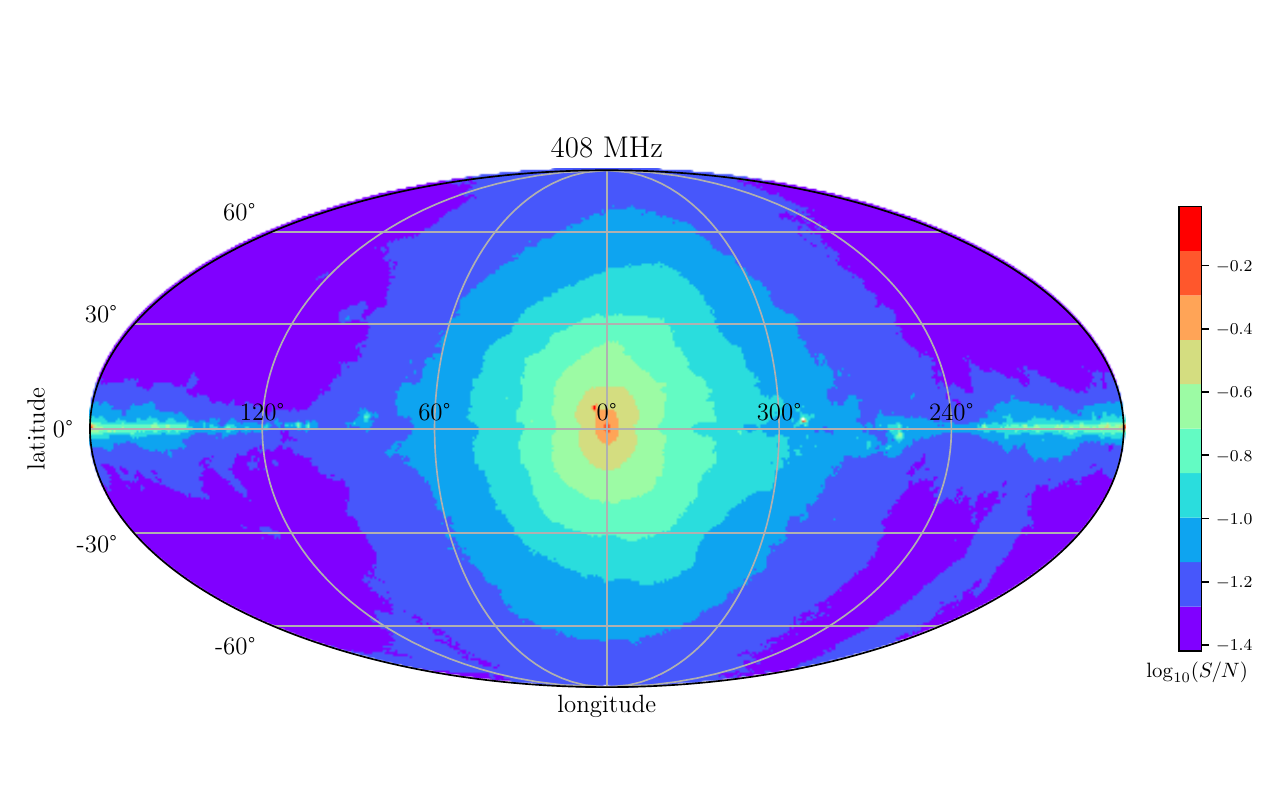}
    \vspace{-60pt}
    
    \includegraphics[width=0.8\textwidth]{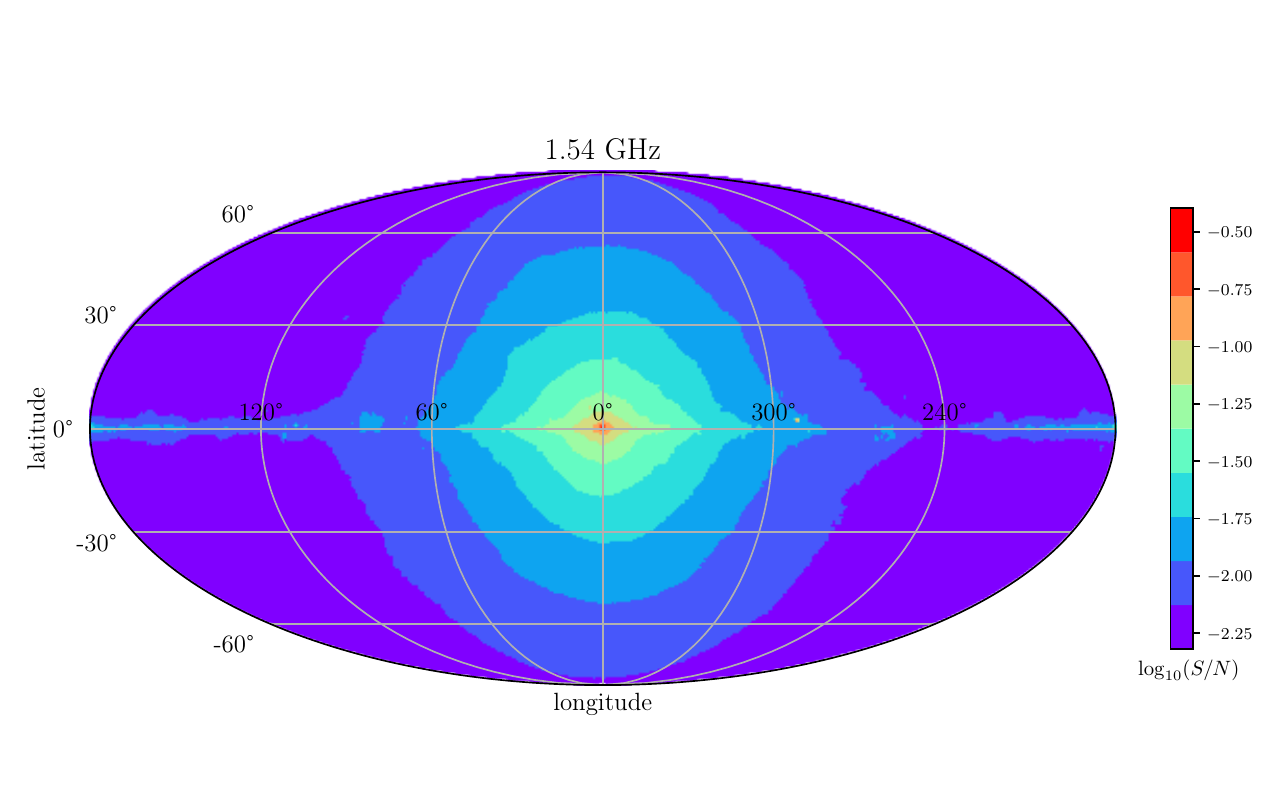}
    \caption{\label{fig: all-sky map of constraints}All-sky map (in Galactic
	coordinates) of the 
	signal-to-noise ratio [cf.~Eq.~\eqref{eqn: signal-to-noise ratio for a telescope}] at $50 \si{\ MHz}$ (top), 
	$408 \si{\ MHz}$ (middle), and $1.54 \si{\ GHz}$ (bottom). 
	The axion-photon coupling is set to 
	$g_{a\gamma} = 10^{-10} \si{\ GeV}^{-1}$ and we have used the NFW profile for DM density distribution. }
\end{figure}

\subsection{Sky maps}
\label{sec:skymaps}
We generated all-sky maps of the expected diffuse flux of radio photons from
stimulated decays of axions at three different frequencies: 
$50 \si{\ MHz}$ (the lowest frequency that SKA-low is sensitive to), 
$408 \si{\ MHz}$ (which matches the frequency of the Haslam map~\cite{haslam1981408,haslam1982408}), 
and $1.54 \si{\ GHz}$ (the highest frequency that SKA-mid can reach). 
Figure \ref{fig: all-sky map of constraints} shows the signal-to-noise ratio
for a fixed $g_{a\gamma} = 10^{-10} \si{\ GeV^{-1}}$  and time-independent Galactic
radio emission. The signal-to-noise ratio is computed from 
Eqs.~(\ref{eqn: signal-to-noise ratio for a telescope}) 
and~(\ref{eqn: signal-to-noise ratio for an array}). 
When performing the integration over $x$ 
in Eq.~(\ref{eqn: signal-to-noise ratio for a telescope}), we take the upper limit
to be the 
virial radius of the Galaxy $R _{\mathrm{vir}}\simeq 221$ kpc [cf.~Section~\ref{subsec: mass density}], neglecting the small
offset of the Earth from the Center of the DM halo. This is a 
good approximation since in Eq.~\eqref{eqn: signal-to-noise ratio for a telescope} the contribution to $S _{\nu}$  from
the edge of the Galactic halo is expected to be insignificant.
Also, when performing the angular integration in Eq.~\eqref{eqn: signal-to-noise ratio for a telescope} we assume that the DM density is constant over each pixel of size 1.7 arcmin, which corresponds to the resolution of the Haslam map.

We compute the optical depth in Eq.~\eqref{eqn: signal-to-noise ratio for a telescope} by setting the upper limit of the spatial integral at  
$x = R _\mathrm{vir}$ in Eq.~(\ref{eqn: optical depth}). The optical depth at higher frequencies turns out to be quite small in all sky directions; therefore, its effect on the sky maps shown in Figure~\ref{fig: all-sky map of constraints} is non-negligible only in the $50 \si{\ MHz}$ case. In order to reduce the computational time, we make an assumption that for a given line-of-sight distance, the optical depth is the same in any sky direction. This approximation results in conservatively estimating the flux of photons, 
especially from decays of axions close to the Earth. On the other hand, for axion decays in the close vicinity of the Galactic Center, Eq.~\eqref{eqn: optical depth} likely underestimates the optical depth. This is because the emission measure 
$\mathrm{EM} = \int\dd x \ n_e^2$ of radio photons from the Galactic Center 
that results from the electron number density in 
Eq.~(\ref{eqn: electron number density}) is 
$\mathcal{O}(10^2) \si{\ pc\ cm^{-6}}$, which is three orders of magnitude smaller than the estimate using the observed radio photon flux from 
Sgr A$^*$~\cite{1989ApJ...342..769P}. 
Nevertheless, we checked that this does not impact the flux from axion decays 
for the following reason. When observing the decay flux in the direction 
of the Galactic Center, the number of photons with momentum pointing
away from the Galactic Center and towards the Earth at an intermediate point $x$
would be $e ^{\tau(x)}$ times larger than that measured in the Haslam map, $f_\gamma(x) = e^{\tau(x)} f_\gamma^\oplus$. This is so because 
only photons unaffected by free-free absorption reach the Earth and are
recorded in the Haslam map. As a result, the factor $e^{\tau(x)}$ cancels out 
the factor $e ^{-\tau(x)}$ in 
Eq.~(\ref{eq:flux_stimulated}), which takes into account the absorption
of photons on their way from the axion decay point to the Earth. 
On the other hand, at the same point $x$, the number of photons traveling in 
the direction of the Galactic Center is $e^{\tau(x)}$ times smaller 
compared to that observed at the Earth, $\tilde{f}_\gamma^{\oplus}$, coming from the Galactic Anti-center direction. Therefore, 
Eq.~(\ref{eq:flux_stimulated}) can be approximated as
\begin{align*}
    \label{eqn: approximated photon flux from axion decays}
    S_{\nu} \simeq  \frac{\Gamma_a}{4\pi\Delta\nu} \int\dd x \int \dd \Omega\ \rho _{a}(x,\Omega) \left(f_{\gamma}(\Omega)+\tilde{f}_{\gamma}(\Omega)e ^{-2\tau(\nu,x,\Omega)} \right) \atag .
\end{align*}
From the Haslam map we deduce that the number of photons from the Galactic 
center $f_{\gamma}(\Omega)$ is about 50 times larger than the number of 
photons from the Anti-Galactic Center $\tilde{f}_{\gamma}(\Omega)$. As a 
result, ${S} _{\nu}$ from the Galactic Center does not depend strongly on the distribution of electrons in the Milky Way. 
This would not be necessarily the case when observing the photon flux from
a direction where fewer background photons are observed compared to 
those from its antipodal direction. 
Nevertheless, we still expect the axion decay flux 
$S _{\nu}$ to be largely independent of the electron 
number density, because for all other directions except near (within $\sim 6^\circ$ of) the Galactic Center, 
the optical depth turns out to be negligible.

Let us stress that Eq.~(\ref{eqn: approximated photon flux from axion decays})
is valid only if two conditions are satisfied: (i) the Galactic radio emission is time-independent, and (ii) the emissivity of photons everywhere along the line of sight is negligibly small, except near the Galactic Center. Therefore, it is fair to say that for diffuse emission, the Galactic Center is the dominant source. 
However, this approximation is clearly not valid for time-dependent point sources like SNRs; therefore, we will use the exact expression for the flux density~\eqref{eq:flux_stimulated} when estimating the gegenschein signal stimulated by the time-dependent radio flux from SNRs in Section~\ref{sec:point}.

\subsection{Galactic Center versus Anti-center}
\label{sec:galc}

From the all-sky maps we deduce that the Galactic Center is the most promising direction to look at for deriving the constraints in the $(m_a,g_{a\gamma})$ plane from diffuse emission of radio photons due to stimulated axion decays. This is reasonable
from Eq.~(\ref{eq:flux_stimulated}) given the large density of axions around 
the Galactic Center, which produces a decay signal large enough to overcome the large background. 
Interestingly, the brightness in the direction of the Galactic Anti-center
also stands out from the sky maps. 
This is so, because this direction benefits from the strong radio 
emission from the Galactic Center that stimulates axion decays and a much
reduced foreground contamination. These two effects compensate for the reduced
density of target DM axions at the Galactic Anti-center. 

\begin{figure}[t!]
    \centering
   \includegraphics[width=0.7\textwidth]{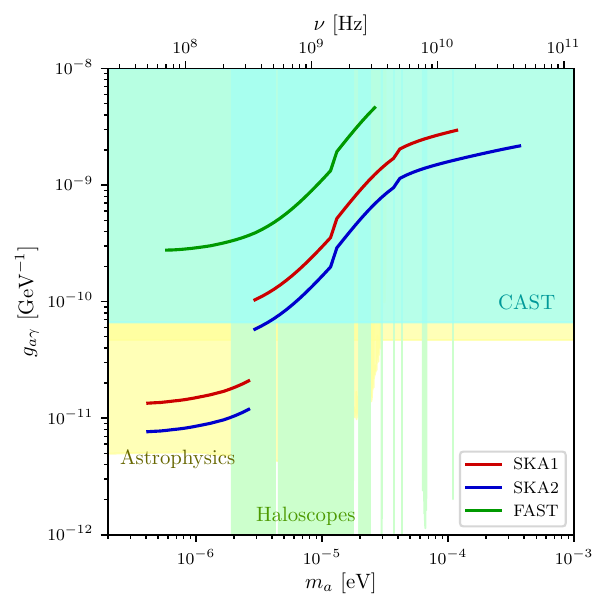}
    \caption{Comparing the sensitivity curves, corresponding to $S/N>1$ in Eq.~\eqref{eqn: signal-to-noise ratio for an array}, for stimulated axion decays in the direction of the Galactic Anti-center, derived using 100 hours of observation time at three different radio telescopes: FAST (green), SKA1 (red) and SKA2 (blue). The discontinuity at 350 MHz for SKA1 and SKA2 is due to the transition from SKA-low to SKA-mid [cf.~Table~\ref{tab: parameters of telescopes}]. The shaded regions show the current exclusion limits from CAST~\cite{CAST:2017uph} (cyan), haloscopes~\cite{ADMX:2021nhd} (light green) and astrophysics~\cite{Dolan:2022kul, Noordhuis:2022ljw} (yellow).}
    \label{fig: constraints_for_each_telescope}
\end{figure}
Before making a comparison of the Galactic Center and Anti-center with the point sources, we would like to compare the sensitivities derived using different radio telescopes. This is shown in Figure~\ref{fig: constraints_for_each_telescope} for 100 hours of observation time in the Galactic Anti-center direction at the three different radio telescopes considered here, namely, FAST (green), SKA1 (red) and SKA2 (blue). Here we have used the NFW profile for the DM density distribution, but other profiles essentially give the same result, because away from the Galactic Center, the variation in the DM density among the different profiles is negligible.   The shaded regions show the existing 95\% C.L. exclusion limits (unless otherwise specified). The helioscope limit from CAST~\cite{CAST:2017uph} is shown by the cyan shaded region.  The green-shaded region shows the collective constraint (as compiled in Ref.~\cite{AxionLimits}) from various haloscope experiments: ADMX~\cite{ADMX:2021nhd}, HAYSTAC~\cite{HAYSTAC:2020kwv, HAYSTAC:2023cam}, ORGAN~\cite{Quiskamp:2022pks}, UPLOAD \cite{Thomson:2019aht}, RBF~\cite{DePanfilis:1987dk, Wuensch:1989sa}, UF~\cite{Hagmann:1990tj}, CAPP \cite{CAPP:2020utb,Lee:2022mnc,Kim:2022hmg,Yi:2022fmn,Yang:2023yry}, CAST-CAPP~\cite{Adair:2022rtw}, QUAX~\cite{Alesini:2020vny,Alesini:2022lnp}, BASE \cite{Devlin:2021fpq}, and TASEH~\cite{TASEH:2022vvu}. The yellow-shaded region is the astrophysical constraint from the $R_2$ parameter~\cite{Dolan:2022kul} (top right, just below CAST) and from pulsar data~\cite{Noordhuis:2022ljw}. We find that the sensitivities derived are comparable to the current constraints, and moreover, can beat the CAST limit in the low-frequency regime. It should also be emphasized that the only constraint that beats our projected sensitivities is the  pulsar limit~\cite{Noordhuis:2022ljw} which is subject to astrophysical modeling uncertainties in the sourced axion spectrum, whereas our stimulated axion decay spectrum induced by diffuse emission is more robust against the astrophysical uncertainties, especially in the Galactic Anti-center direction. Since SKA2, not surprisingly, gives us the best sensitivity, we will show our following results only for SKA2.   

\begin{figure}[t!]
    \centering
    \includegraphics[width=0.7\textwidth]{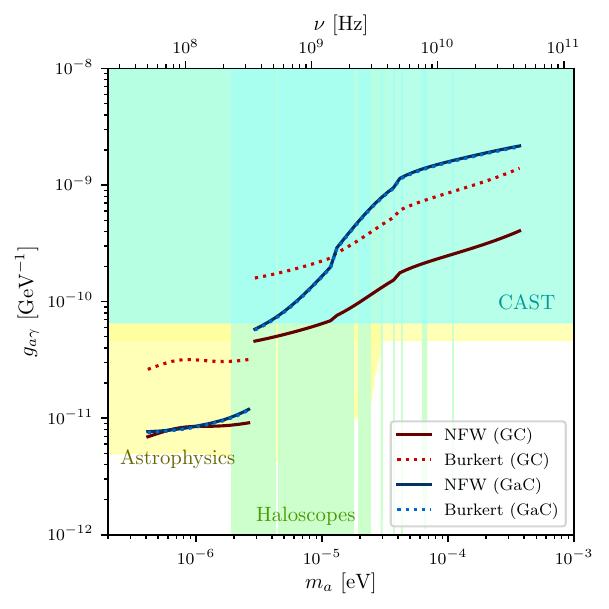}
    \caption{SKA2 sensitivity curves, corresponding to $S/N>1$ in Eq.~\eqref{eqn: signal-to-noise ratio for an array}, for stimulated axion decays in the direction of Galactic Center (dark red) and Galactic Anti-center (blue), obtained using NFW (solid) and Burkert (dotted)  profiles for the DM density distribution in the Galactic halo. The shaded regions show the current exclusion limits from CAST~\cite{CAST:2017uph} (cyan), haloscopes~\cite{ADMX:2021nhd} (light green) and astrophysics~\cite{Dolan:2022kul, Noordhuis:2022ljw} (yellow).}
    \label{fig: constraints1}
\end{figure}

In Figure~\ref{fig: constraints1}, we compare the results in the direction of the Galactic Center (dark red) and Galactic Anti-center (blue). The solid (dotted) lines are obtained using NFW (Burkert) profile for the DM density distribution in the Galactic halo. Again we have used 100 hours of observation time at SKA2 and $S/N>1$ to derive the sensitivity curves. The Galactic radio emission is assumed to be constant in time. The shaded regions show the current exclusion, as in Figure~\ref{fig: constraints_for_each_telescope}. 

The constraints from the Galactic Center can be compared to those reported 
in Ref.~\cite{Caputo:2018vmy}. To reiterate the differences, in our analysis we include an amplification factor 
$e^{\tau(x)}$ in the photon emissivity and we use the Haslam $408 \si{\ MHz}$ map 
to describe the Galactic radio emission, while Ref.~\cite{Caputo:2018vmy}
used the observed photon energy density from the Galactic Center 
at $\nu = 1.4 \si{\ GHz}$ from Ref.~\cite{Yusef-Zadeh:2012efm}. In fact, their $f_\gamma$ from the Galactic Center is 7-8 times larger than ours, but since it increases both the signal flux and the background by the same rate, the final result is similar to ours. In any case, the Galactic Anti-center result presented here is our main new point.

Possible uncertainties in our results could come from our limited knowledge
of the electron distribution model, the axion DM distribution, and 
the spectrum of Galactic radio photon emission. 
The electron density enters the calculation of the free-free absorption of 
radio photons. This is an important effect, particularly at lower frequencies 
corresponding to $m _{a} \lesssim 10 ^{-6} \si{\ eV}$. The constraints from 
the Galactic Center are more susceptible to this absorption because of the 
large electron number density around the Galactic Center. The same goes for 
the density of DM axions in the central regions of the Galaxy, 
which can vary by orders of magnitude depending on whether one assumes
a cuspy or a cored profile.

In contrast, the signal from the Galactic Anti-center is more robust, since
there is less variation in the predictions for the mass density of DM axions in the outer parts of the Galaxy. Thus, the strength of these 
constraints will remain intact even assuming that the DM profile is
cored, as we can see clearly in Figure \ref{fig: constraints1}. Moreover, 
the constraints from the Galactic Center might be weakened at low-frequencies,
since, as mentioned before, we might be underestimating
the emission measure EM in the direction of the Galactic 
Center~\cite{1989ApJ...342..769P}. In the frequency region that we consider, 
the number of photons is nearly proportional to the brightness temperature: 
$f _{\gamma} = (e ^{E_\gamma/k_B T}-1)^{-1} \simeq k_B T/E_\gamma$. Thus, the 
signal-to-noise ratio is not sensitive to the frequency-dependence of the 
brightness temperature of the Galactic radio emission, as long as its 
contribution to the signal and the noise are dominant, as is the case 
at lower frequencies. Therefore, the uncertainty in the spectrum of Galactic 
radio photons will mainly affect the constraints at higher frequencies. 

\subsection{Point source sensitivity}
\label{sec:point}
\begin{figure}[t!]
    \centering
   \includegraphics[width=0.99\textwidth]{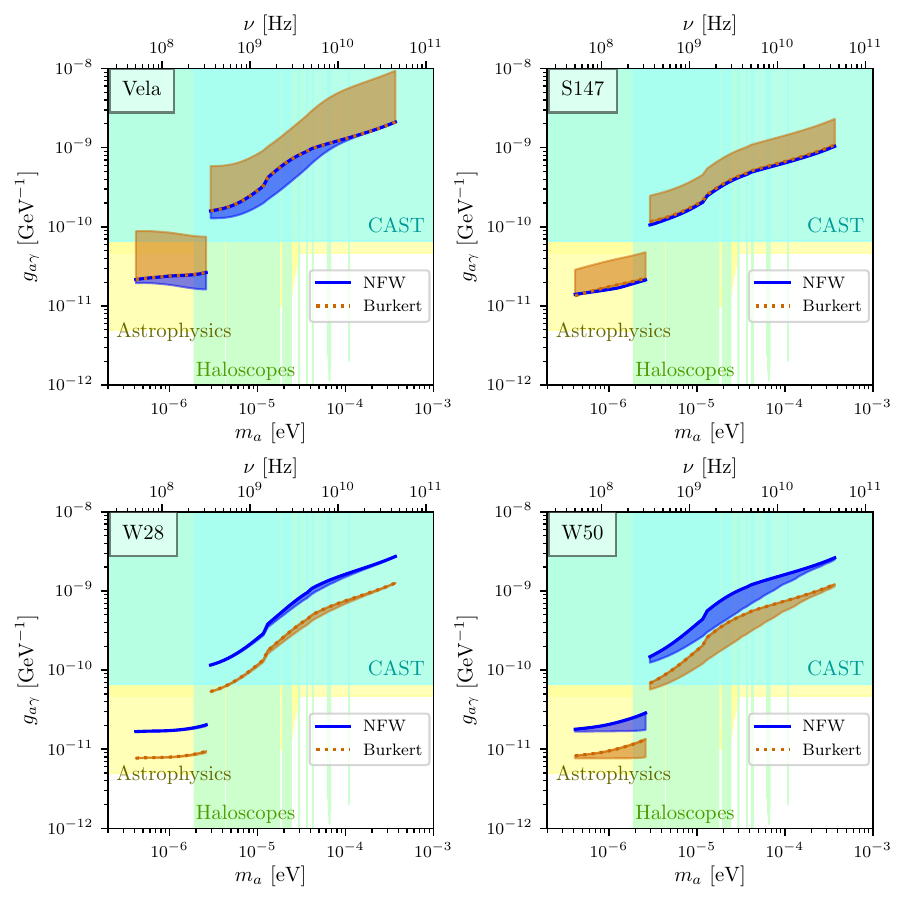}
 \caption{SKA2 sensitivity curves for the gegenschien signal in the direction opposite to the SNR sources Vela (top left), S147 (top right), W28 (bottom left) and W50 (bottom right). The solid (dotted) curves correspond to the NFW (Burkert) profiles for the DM density distribution in the Galactic halo. The bands associated with each curve takes into account the SNR parameter uncertainties, as discussed in Section~\ref{subsec: radio emission from SNRs}. The shaded regions show the current exclusion limits from CAST~\cite{CAST:2017uph} (cyan), haloscopes~\cite{ADMX:2021nhd} (light green) and astrophysics~\cite{Dolan:2022kul, Noordhuis:2022ljw} (yellow).}
  \label{fig:SNR}
\end{figure}

Now we consider the four point sources listed in Table~\ref{tab: snr parameters} and calculate the gegenschien signal from each of them. Figure~\ref{fig:SNR} summarizes the constraints obtained by observing in the opposite directions to these SNRs. 
We first extract the contribution from Galactic synchrotron radiation 
from the Haslam map by subtracting the flux from supernova remnants estimated 
by extrapolating their observed flux with the spectral indices listed in 
Table~\ref{tab: snr parameters}. The image of an SNR is approximated by a 
circle with a diameter given in Table~\ref{tab: snr parameters}, and it is 
assumed to be unchanged during the evolution of the SNR. 
As before, we assume that the Galactic radio emission is
time-independent. The time 
and frequency dependence of the radio emission from supernova remnants is
estimated following the discussion in Section~\ref{subsec: radio emission from SNRs} with
the spectral index for each SNR as listed in Table~\ref{tab: snr parameters}. Then, SKA2 sensitivities are 
obtained by letting $S/N = 1$ in 
Eq.~(\ref{eqn: signal-to-noise ratio for an array}). The bands in Figure~\ref{fig:SNR} capture the uncertainties associated with the SNR parameters, as discussed in Section~\ref{subsec: radio emission from SNRs}. The dominant uncertainties come from the modeling of the flux and the MFA time. We have chosen the conservative model (ii) $S_\nu\propto t^{-2(\gamma+1)/5}$ and $t_{\rm MFA}=100$ years to draw the central curves and the bands are obtained by varying $t_{\rm MFA}$ between 30 and 300 years, as well as by using model (i) $S_\nu\propto t^{-4\gamma/5}$. The other SNR parameter uncertainties listed in Table~\ref{tab: snr parameters} are also included, although their effects on the total uncertainty is small, except for the SNR age.

To reduce the computational time, two simplifications are made when 
estimating the results for W28, W50, and Vela. 
Firstly, free-free absorption is neglected in the frequency band measured
by SKA-mid. 
We checked that this results in a negligible shift due to the small optical 
depth $\tau \simeq 10^{-3}$. On the other hand, the effect of free-free 
absorption is taken into account at the frequencies
corresponding to SKA-low using the approximation 
$\tau \simeq \tau(\nu,x=R _{\mathrm{vir}},\Omega=(l,b))$. That is, 
we calculate $\tau$ only in the opposite direction of each SNR and 
we neglect the directional-dependence of $\tau$ inside the instantaneous 
field of view. This approximation is not valid for S147, because
its anti-direction is close to the Galactic Center, and the optical depth is 
expected to depend strongly on the direction.

The strongest constraints at high frequencies are placed by S147 mainly because of the large axion mass density at the Galactic Center. The kink at 
$m_a \simeq 15 \si{\ eV}$ is caused by the abrupt change in the spectral 
index of S147 from $\alpha = 0.3$ to 1.2 at $\nu= 1.7 \si{\ GHz}$. A lower DM mass density at the Galactic Center would weaken the 
constraints from the Galactic Center and from S147. 
One mechanism that could remove mass from the inner region of the Galaxy 
is mass segregation due to dynamical friction.  However, as discussed in Section~\ref{subsec: gravitational interactions of axions}, this effect turns out to be negligible. Comparing our results for Vela, W28 and W50 with those in Refs.~\cite{Buen-Abad:2021qvj,Sun:2021oqp}, we see that their results obtained from observations in the interferometric mode are better at high frequencies, whereas our results derived in the single-dish mode are better at low frequencies, where we can beat the CAST limit and be competitive with the pulsar limit.

\begin{figure}[t!]
    \centering
    \includegraphics[width=0.99\textwidth]{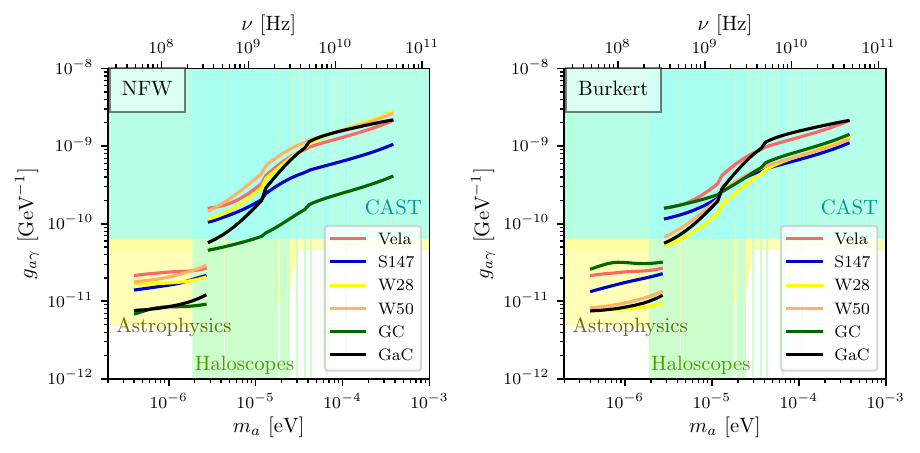}
    \caption{The summary plot comparing the SKA2 sensitivities for the stimulated axion decay signal for all the sources considered in this study, namely, the four SNRs, as well as the diffuse emission from the Galactic Center and Anti-center. The left (right) panel is assuming NFW (Burkert) profile for the DM density distribution in the Galactic halo. For the SNRs, we only show the central values of the sensitivities. }
    \label{fig:summary}
\end{figure}

Finally, in Figure~\ref{fig:summary}, we summarize our results by collecting the sensitivities from all six sources considered in this study, namely, the gegenschein signals from the four SNRs shown in Figure~\ref{fig:SNR}, as well as the stimulated decay signals from the diffuse flux in the directions of Galactic Center and Anti-center shown in Figure~\ref{fig: constraints1}. We find that at high frequencies ($\nu>350$ MHz), corresponding to $m_a>3.5\times 10^{-6}$ eV, the Galactic Center gives the best sensitivity for the NFW profile. On the other hand, at low frequencies ($\nu<350$ MHz), the Galactic Anti-center gives better sensitivity than the Galactic Center, irrespective of the density profile. The point source sensitivities are better than the diffuse ones only for a cored profile like the Burkert.    

\section{Conclusions}
\label{sec: conclusions}

We have re-examined the constraints on the axion-photon coupling placed by
the non-observation of radio photons from the stimulated decay of
axion dark matter particles. In addition to the constant 
extragalactic radio emission and the CMB, we have included the Galactic radio 
emission with the spatial density empirically determined from the Haslam map 
at $408$ MHz. With these diffuse sources of radio emission through the
DM halo, we 
proceed to generate a full-sky map of the expected signal assuming $100$ hours
of observation each with the FAST and SKA telescopes. We also consider four selected 
supernova remnants that could potentially generate bright counterimages.
Our final results are displayed in Figure~\ref{fig:summary} and show that,
if the DM density follows a cuspy NFW profile, the Galactic Center
provides the strongest constraints of order $g _{a\gamma} \lesssim$ a few 
$10 ^{-11} \si{\ GeV ^{-1}}$ for the frequencies observed by the 
SKA-low telescope. 

Interestingly, our full-sky maps also suggest that 
the direction of the Galactic Anti-center can be used to place competitive
constraints. Unlike the direct emission from the Galactic Center,
the estimates from the Anti-center (which is technically the 
{\itshape gegenschein emission from the center})
do not depend on the highly uncertain DM density at the center of 
the halo, nor are impacted by a large optical depth. They are also arguably 
more robust than the limits from supernov\ae\ that depend on assumptions about 
the radio emission during early stages in their evolution. 
Furthermore, if the Galactic DM halo is described by a cored
profile, {\itshape the anti-center 
places not only the most robust but the most stringent constraints on the
axion-photon coupling}.

\medskip 
\noindent
\textbf{Note added:} During the final stages of our work, 
Refs.~\cite{Sun:2023gic, Todarello:2023xuf} appeared. In~\cite{Sun:2023gic}, the authors update their 
previous analysis of the gegenschein signal and also consider the 
direct emission (``forwardschein"). Our conclusions largely agree 
where we overlap. Nevertheless, their focus is on point sources, while we 
compare the point source sensitivity to the diffuse emission sensitivities from the Galactic Center and 
Anti-center. In Ref.~\cite{ Todarello:2023xuf},  the authors provide a detailed and general derivation of the effects, and focused on the case of Galactic pulsars as stimulating sources. The ensuing sensitivities are comparable to the SNR sensitivities shown here.

\section*{Acknowledgments}
We thank Sam Witte for useful discussion. This work was partly supported by the U.S. Department of Energy under grant No.~DE-SC 0017987. 
\appendix

\section{Stimulated decay of axions into photons}
\label{sec:axion_decay_details}
In this section we review the derivation of the stimulated axion decay in a photon bath,
which was first discussed 
in Refs.~\cite{Tkachev:1986tr,Tkachev:1987cd,Kephart:1994uy} as 
a possible energy source in the central parts of dense axion clusters. 
The stimulated decay of axions is induced by the interaction term given in Eq.~\eqref{eq:lag} and is greatly enhanced in astrophysical environments in the presence of an ambient radiation field and a
large axion number density.
We consider the general case where the
photon distribution functions are not necessarily spherically symmetric (as assumed in Ref.~\cite{Buen-Abad:2021qvj}).

To find the evolution of the phase-space distribution function of photons
$f_\gamma (\mathbf{x},\mathbf{p},t) \equiv f_\lambda$, we consider the process 
$a\pqty{\mathbf{p}_a}\rightarrow \gamma \pqty{\mathbf{p}_{1};\lambda}+
\gamma \pqty{\mathbf{p}_{2};\lambda}$ and its inverse 
$\gamma\left(\mathbf{p}_{1};\lambda\right) + 
\gamma\left(\mathbf{p}_{2};\lambda\right) \rightarrow 
a\left(\mathbf{p}_{a}\right)$, where $\lambda = +,-$ denotes the helicity of photons. The conservation of angular momentum requires 
that the two photons emitted from the decay of a scalar or pseudo-scalar 
particle should have the same helicity, and the same applies
to the two incoming photons in the inverse process. 
The evolution of the phase-space distribution function of the photon with
momentum $\mathbf{p}_1$ is given by the 
Boltzmann 
equation~\cite{Tkachev:1986tr,Tkachev:1987cd,Kephart:1994uy,Buen-Abad:2021qvj}
\begin{align*}
    \frac{\mathrm{d}}{\mathrm{d} t} f_{1\lambda}
    &= \frac{1}{2E _{1}} \int \frac{\dd ^{3} p _{a}}{(2\pi) ^{3}2 E _{a}} \int \frac{\dd ^{3} p _{2}}{(2\pi) ^{3} 2E _{2}} |\mathcal{M}
\left(f_{a}\left(1+f_{1\lambda}+f_{2\lambda}\right)-f_{1\lambda} f_{2\lambda}\right) (2 \pi)^{4}\delta^{(4)}\left(p_{a}-p_{1}-p_{2}\right), \atag
\end{align*}
where $\mathcal{M} \equiv \mathcal{M}(a\left(\mathbf{p}_{\mathbf{a}}\right) \rightarrow \gamma\left(\mathbf{p}_{1};+\right)+\gamma\left(\mathbf{p}_{2};+\right)) = \mathcal{M}(a\left(\mathbf{p}_{\mathbf{a}}\right) \rightarrow \gamma\left(\mathbf{p}_{1};-\right)+\gamma\left(\mathbf{p}_{2};-\right))$ 
is the invariant decay amplitude determined by the Abelian chiral
anomaly~\cite{Adler:1969gk}.

For all the environments that we consider the phase-space density of axions is
much larger than that of photons, $f_a \gg f_{i\lambda}$, so we can neglect
the last term $\propto f _{1} f _{2}$ corresponding to the inverse decay
process in the collision integral. The first term $\propto f_a$  describes 
the spontaneous decay of axions which, as shown in Eq.~\eqref{eq:vdecay}, has too large a
lifetime to be experimentally detectable, and thus we also neglect it.
The second and third terms, $\propto f_a f_{i\lambda}$ correspond to the 
stimulated production of photons.

Assuming a parity symmetric photon field, $f_{i+} = f _{i-} = 
\frac{1}{2}f _{\gamma}$, we can sum over the helicities to obtain:
\begin{align*}
    \label{eqn: boltzmann equation}
    \frac{\mathrm{d}}{\mathrm{d} t} f_{1} =
    \frac{\mathrm{d}}{\mathrm{d} t}  \sum _{\lambda} f_{1\lambda} 
    &
   = \frac{1}{2E _{1}} \int \frac{\dd ^{3} p _{a}}{(2\pi) ^{3}2 E _{a}}  \int \frac{\dd ^{3} p _{2}}{(2\pi) ^{3} 2E _{2}}  (2 \pi)^{4}\delta^{(4)}\left(p_{a}-p_{1}-p_{2}\right) \\
    &\ \ \ \ \ \times f_{a}\left(f_{1}+f_{2}\right) \frac{1}{2} \sum _{\lambda} |\mathcal{M}(a\left(\mathbf{p}_{\mathbf{a}}\right) \rightarrow \gamma\left(\mathbf{p}_{1};\lambda\right)+\gamma\left(\mathbf{p}_{2};\lambda\right))| ^{2}.   \atag
\end{align*}

To leading order we can take the axion dark matter to be cold, neglecting
the small velocity dispersion in the Galaxy $\sigma_{v} \approx 5 \times 10^{-4}$ \cite{Freese:2012xd}. With this assumption, the axion distribution 
function is given by $f_{a}=(2 \pi)^{3} n_{a}(\mathbf{x})\ 
\delta^{(3)}\left(\mathbf{p}_{a}\right)$, where $n_{a}(\mathbf{x})$ is the axion number density. Thus, the photon production 
rate is 
\begin{align*}
    \frac{\mathrm{d}}{\mathrm{d} t} f_{1}
    &= \frac{\rho_a(\mathbf{x})}{4 m_a^2 E_1} \int 
	\frac{\dd ^{3} p _{2}}{(2\pi) ^{3} 2E_{2}} 
	\left(f_{1}(\mathbf{x} ,\mathbf{p _{1}}, t)+f_{2}(\mathbf{x} ,
	\mathbf{p _{2}}, t) \right)(2 \pi)^{4}\delta\left(m_{a}-E_{1}-
	E_{2}\right) \delta^{(3)}\left( \mathbf{p _{1}} + 
	\mathbf{p _{2}} \right) 
 \\
    &\ \ \ \ \ \ \ \ \ \ \ \ \ \times 
    \frac{1}{2} \sum _{\lambda} |\mathcal{M}(a\left(\mathbf{0}\right) \rightarrow \gamma\left(\mathbf{p}_{1};\lambda\right)+\gamma\left(\mathbf{p}_{2};\lambda\right))| ^{2} \atag ,
\end{align*}
where $\rho _{a} (\mathbf{x}) = m _{a} n _{a}(\mathbf{x})$. 
Integrating over $\mathbf{p_2}$ and identifying $f_i$ with $f_\gamma$ (traveling either forward or backward along the line of sight), we get
\begin{align*}
    \label{eqn: boltzmann equation 2}
    \frac{\mathrm{d}}{\mathrm{d} t} f_{\gamma}(\mathbf{x} ,\mathbf{p}, t) &= \frac{\pi \rho _{a}(\mathbf{x})}{4 m_a^2 E_\gamma^2} \left(f_{\gamma}(\mathbf{x} ,\mathbf{p}, t)+f_{\gamma}(\mathbf{x} , -\mathbf{p}, t) \right) \delta\left(m_{a}-2E_\gamma\right) 
    \frac{1}{2} \sum _{\lambda} |\mathcal{M}(a\left(\mathbf{0}\right) \rightarrow \gamma\left(\mathbf{p};\lambda\right)+\gamma\left(-\mathbf{p};\lambda\right))| ^{2} \atag .
\end{align*}
We see that the energy $E$ of the two outgoing photons is equal to $m_a/2$ due
to the conservation of energy. Substituting the expression of the spontaneous
decay rate in the rest frame of an axion, 
\begin{align*}
    \label{eqn: spontaneous decay rate of axion}
\Gamma _{a} = \tau _{a} ^{-1}= 
	\frac{m_{a}^{3}g_{a \gamma}^{2}}{64\pi}= \frac{1}{8\pi} \frac{1}{2m _{a}}
\frac{1}{2} \sum _{\lambda} |\mathcal{M}(a\left(\mathbf{0}\right) \rightarrow \gamma\left(\mathbf{p};\lambda\right)+\gamma\left(-\mathbf{p};\lambda\right))| ^{2}, \atag
\end{align*}
into Eq.~(\ref{eqn: boltzmann equation 2}), we finally obtain the time derivative of the photon distribution function as
\begin{align}
    \frac{\mathrm{d}}{\mathrm{d} t} f_{\gamma}(\mathbf{x} ,\mathbf{p}, t) =
	\frac{\pi^2 \Gamma _{a} \rho _{a}(\mathbf{x})}{E_\gamma^3} 
	\left(f_{\gamma}(\mathbf{x} ,\mathbf{p}, t)+f_{\gamma}(\mathbf{x} , 
	-\mathbf{p}, t) \right) \delta\left(E_\gamma -m_{a}/2\right).
 \label{eq:dist}
\end{align}
This expression implies that we will observe photons from decays of axions 
stimulated by the photons traveling either towards the Earth (the first term) 
or away from the Earth (the second term) and both having energy equal to 
half of the axion mass. Although we assumed that the axion DM is
at rest in the Galaxy, this conclusion is true in general: two photons with 
the same energy are emitted back-to-back in the rest frame of the axion.
\\
\indent
Integrating the photon distribution function obtained from Eq.~\eqref{eq:dist} over the phase space solid angle, we get the phase space density $S_\nu = 2\nu^3 \int d\Omega f_\gamma$, where $\nu = E_\gamma/2\pi=m_a/4\pi$ is the frequency of the photon. Integrating the phase space density over the distance along the line of sight, and averaging over the photon bandwidth, we obtain the flux density given in Eq.~\eqref{eq:flux_stimulated}.

\bibliographystyle{utcaps_mod} 

\bibliography{main}

\end{document}